\documentclass[journal]{IEEEtran}
\usepackage{graphics}
\usepackage{mathrsfs}
\usepackage{color}
\usepackage{amsfonts} 
\usepackage{color} 
\usepackage{float}
\usepackage{cite}
\usepackage{graphicx}
\usepackage{amsmath}
\usepackage{amsthm}
\usepackage{multicol}
\usepackage{multirow}
\usepackage{color}
\usepackage{ifpdf}
\usepackage{flushend}
\usepackage{txfonts}
\usepackage{mdwlist}

\usepackage{CJK}

\theoremstyle{definition} \newtheorem{remark}{Remark}
\theoremstyle{definition} 
\theoremstyle{definition} \newtheorem{theorem}{Theorem}
\theoremstyle{definition} \newtheorem{lemma}{Lemma}
\theoremstyle{assumption} 
\theoremstyle{definition} \newtheorem{definition}{Definition}
\theoremstyle{definition} \newtheorem{claim}{Claim}
\theoremstyle{definition} 
\theoremstyle{definition} \newtheorem{corollary}{Corollary}
\theoremstyle{definition} 

\usepackage{ifpdf}
\ifpdf
  \usepackage{epstopdf}
\fi

\newcommand{\R}{\mathbb{R}}

\newcommand{\C}{\mathcal{C}}
\newcommand{\setS}{\mathcal{S}}

\newcommand{\X}{\mathcal{X}}
\newcommand{\Id}{\text{Id}}

\begin{document}

\title{Small-Gain Theorem for Safety Verification under High-Relative-Degree Constraints}

\author{Ziliang~Lyu,~
        Xiangru~Xu,~
        and~Yiguang~Hong

\thanks{Z. Lyu (e-mail: ziliang\_lyu@outlook.com) and Y. Hong (e-mail: yghong@iss.ac.cn) are with the Department of Control Science and Engineering, Tongji University, Shanghai, China}

\thanks{X. Xu (e-mail: xiangru.xu@wisc.edu) is with the Department of Mechanical Engineering, University of Wisconsin-Madison, Madison, WI, USA.}
}

\maketitle

\begin{abstract}
This paper develops a small-gain technique for the safety analysis and verification of interconnected systems with high-relative-degree safety constraints. In this technique, input-to-state safety (ISSf) is used to characterize how the safety of a subsystem is influenced by the external input, and ISSf-barrier functions (ISSf-BFs) with high relative degree are employed to capture the safety of subsystems. With a coordination transform, the relationship between ISSf-BFs and the existing high-relative-degree (or high-order) barrier functions is established in order to simplify the ISSf analysis. With the help of high-relative-degree ISSf-BFs, a small-gain theorem is proposed for safety verification. It is shown that, under the small-gain condition, i) the interconnection of ISSf subsystems is still ISSf; and ii) the overall interconnected system is input-to-state stable (ISS) with respect to the compositional safe set. The effectiveness of the proposed small-gain theorem is illustrated on the output-constrained decentralized control of two inverted pendulums connected by a spring mounted on two carts.
\end{abstract}


\section{Introduction}

Safety is a fundamental property of practical control systems, e.g., air traffic management systems \cite{mitchell2005time}, industrial robots \cite{vasic2013safety}, life support devices \cite{glavaski2005safety} and autonomous vehicles \cite{ames2016control}. Ensuring safety is important for these safety-critical systems. Over the past years, a set of approaches have been developed for safety verification, including model checking \cite{kupferman2001model}, barrier approaches \cite{prajna2007framework, ames2016control}, and reachability analysis \cite{mitchell2005time}.

Barrier functions become popular because they verify safety with Lyapunov-like arguments, and help avoid the computation of abstractions or reachable sets. A computational method was developed in \cite{prajna2007framework} to search for a barrier function via the sum-of-squares (SOS) optimization. A less conservative barrier function, called the zeroing barrier function (ZBF), was proposed in \cite{xu2015robustness}, where the state trajectories are allowed to get close to the boundary of the safe set, and was extended to the case with arbitrary relative degree in \cite{nguyen2016exponential, xu2018constrained, tan2021high, xiao2021high}. However, it is difficult for high-dimensional systems to synthesize a barrier function. In fact, the computational cost of finding a polynomial barrier function via sum of squares optimization grows polynomially with respect to the system dimension for fixed polynomial degrees, as indicated in \cite{prajna2007framework}. Since a complicated system is often the interconnection of subsystems, a feasible approach is to construct barrier functions for the subsystems individually and then compose them to establish safety for the overall interconnected system.

Small-gain technique is a fundamental tool for the analysis of interconnected systems. The classical small-gain theorem, pioneered by \cite{zames1966input, sandberg1964frequency, desoer2009feedback}, was originally established from the input-output viewpoint with linear gains. A generalization of the small-gain theorem was presented in \cite{hill1991generalization} for feedback interconnections with nonlinear gains. In \cite{Jiang1994Small, jiang1999combined, teel1996nonlinear}, the nonlinear small-gain theorem was developed with help of the input-to-state stability (ISS) framework \cite{sontag1989smooth}. More recently, the ISS small-gain theorem has been generalized to switched systems \cite{long2013small}, hybrid systems \cite{liberzon2014lyapunov} and large-scale networks \cite{dashkovskiy2010small}. Also, the small-gain theorem is useful in various control designs, such as adaptive control \cite{middleton1988design} and event-triggered control \cite{liu2015small}.

Input-to-state safety (ISSf) \cite{kolathaya2018input} is the counterpart of ISS in safety analysis. This concept has attracted a lot of attention in the control community. In \cite{lyu2022small, kolathaya2018input}, two ISSf-barrier functions (ISSf-BFs) were proposed to establish ISSf. The equivalence of these ISSf-BFs has been shown in \cite{lyu2022small}. The ISSf-BFs have been used in the recent paper \cite{krstic2021inverse} to design an inverse optimal safety-critical controller. Even though the small-gain theorem is important for system analysis, there are few results in safety verification, except for \cite{lyu2022small, jagtap2020compositional}, where the individual ISSf-BFs have relative degree one. However, there are many practical systems with high-relative-degree safety constraints, such as Euler-Largrange systems.

The objective of this paper is to provide a small-gain framework for safety analysis and verification when the relative degree of safety constraints is larger than one. Compared with the relative-degree-one results \cite{lyu2022small, jagtap2020compositional}, the main difficulty lies in the high-order derivatives involved in the individual ISSf-BFs and the construction of a compositional ISSf-BF for the interconnected systems. We focus on two small-gain fundamental problems:
\begin{itemize}
  \item the relationship between the ISSf-BFs and the high-order ZBFs of \cite{tan2021high};
  \item the sufficient condition for the ISSf of interconnected systems under high-relative-degree safety constraints.
\end{itemize}

The main contribution of this paper can be summarized as follows.
\begin{itemize}
  \item We develop a new ISSf approach to understand the dependence of the safety on the magnitude of external inputs under high-relative-degree safety constraints. In fact, the ISSf-BF can be converted into an auxiliary ZBF with the help of a coordinate transformation. In this way, we can establish the ISSf by analyzing the auxiliary ZBF condition. This analysis also provides new insight for the ISSf verification.
  \item We propose a small-gain theory for safety analysis and verification under high-relative-degree safety constraints. We show that, with our small-gain condition, i) the interconnection of ISSf subsystems is still ISSf; and ii) the interconnected system is ISS with respect to the compositional safe region. Thus, for the case without external inputs, the interconnected system is always safe if it is initialized safely, and moreover, it becomes safe eventually even though it is initialized outside the safe region.
  \item We develop a comparison technique to prove our main result from an input-output viewpoint.  Different from the existing ideas by analyzing the high-order derivatives of individual ISSf-BFs or constructing a compositional one, we focus on how the individual ISSf-BF is influenced by the external inputs and the initial values of its high-order derivatives. A strong point of this technique is that it does not require any forward completeness assumptions.
\end{itemize}

The remainder of this paper is organized as follows. In Section \ref{Sec:ISSf-BF}, we provide a sufficient condition based on barrier functions for establishing ISSf under high-relative-degree safety constraints. Then a small-gain theorem is developed in Section \ref{sec:small-gain} for the ISSf analysis and verification of feedback interconnections of ISSf subsystems. The effectiveness of this result is illustrated in Section \ref{sec:example} with the decentralized control of two inverted pendulums connected by a spring mounted on two carts (shorted as the pendulum-spring-cart system) with output constraints. Finally, we summarize the conclusions in Section \ref{sec:concl}.

\textbf{Notations.} Throughout this paper, `$\circ$' denotes the composition operator, i.e., $f\circ{g}(s)=f(g(s))$; `T' denotes the transpose operator; $\alpha'(s)$ denotes the derivative of a continuously differentiable function $\alpha$ at $s$; $\R$ and $\R_{\geq0}$ denote the set of real numbers and nonnegative real numbers, respectively. For any $x$ in Euclidean space, $|x|$ is its norm, and $|x|_{\setS}=\inf_{s\in\setS}|x-s|$ denotes the point-to-set distance from $x$ to the set $\setS$. Denote by $L_\infty^m$ the set of essentially bounded measurable functions $u:\R_{\geq0}\rightarrow\R^m$. For any $u\in L_\infty^m$, $\|u\|_J$ stands for the supremum norm of $u$ on an interval $J\subseteq\R_{\geq0}$ (i.e., $\|u\|_{J}=\sup_{t\in J}|u(t)|$), and we take $\|u\|=\|u\|_{[0,\infty)}$ for simplicity. A continuous function $\gamma$: $\mathbb{R}_{\geq0}\rightarrow\mathbb{R}_{\geq0}$ with $\gamma(0)=0$ is of class $K$, if it is strictly increasing. A class $K$ function $\gamma$ is of class $K_\infty$ if it is unbounded. A function $\beta:\R_{\geq0}\times\R_{\geq0}\rightarrow\R_{\geq0}$ is of class $KL$, if for each fixed $t$, the mapping $s\mapsto\beta(s,t)$ is of class $K$, and for each fixed $s\geq0$, $t\mapsto\beta(s,t)$ is decreasing to zero as $t\rightarrow+\infty$. Since barrier functions do not have the positive definiteness of Lyapunov functions, we introduce the following extended comparison functions accordingly. A continuous function  $\gamma:\R\rightarrow\R$ with $\gamma(0)=0$ is of extended class $K$ if it is strictly increasing. In particular, an extended class $K$ function $\gamma$ is of extended class $K_\infty$ if it is unbounded. A function $\beta:\R\times\R_{\geq0}\rightarrow\R$ is of extended class $KL$, if for each fixed $t$, the mapping $s\mapsto\beta(s,t)$ is of extended class $K$, and for fixed $s>0$ and $s<0$, $t\mapsto\beta(s,t)$ is decreasing and increasing to zero, respectively, as $t\rightarrow+\infty$.


\section{Input-to-State Safety Under High-Relative-Degree Safety Constraints}\label{Sec:ISSf-BF}

This section provides a sufficient condition based on barrier functions for ISSf under high-relative-degree safety constraints.

\subsection{Input-to-State Safety}

Consider the system
\begin{flalign}\label{eq:system-with-input}
    \dot{x}=f(x,u),\;\;x(0)=x_0
\end{flalign}
where $x\in\R^{n}$ is the state, $u\in L_\infty^m$ is the external input (maybe ``control'' or ``disturbance'' of the system), and $f:\R^n\rightarrow\R^n$ is locally Lipschitz. For any $x_0\in\R^{n}$ and $u\in L_\infty^m$, the solution of (\ref{eq:system-with-input}), defined on some maximal interval $I(x_0,u)$, is denoted by $x(t,x_0,u)$ (and sometimes by $x(t)$ for simplicity if there is no ambiguity). System (\ref{eq:system-with-input}) is said to be forward complete if $I(x_0,u)=\R_{\geq0}$.

Suppose that the safety constraints of system (\ref{eq:system-with-input}) are characterized by the closed set
\begin{flalign}\label{eq:single-sys-constraint-set}
    \setS_0=\{x\in\R^n:h(x)\geq0\}
\end{flalign}
where $h:\R^n\rightarrow\R$ is a sufficiently differentiable function. Define a larger set
\begin{flalign}\label{eq:ISSfset}
        \C_0=\{x\in\R^n:h(x)+\gamma(\|u\|)\geq0\}
\end{flalign}
where $\gamma$ is a class $K_\infty$ function. We say that $\C_0$ is robustly forward invariant (c.f. \cite[Def. 4.3]{blanchini2008set}), if for all $x_0\in\C_0$ and any $u\in L_\infty^n$, $x(t,x_0,u)\in\C_0$ for all $t\in I(x_0,u)$.

This paper concentrates on the situation when $h$ has relative degree $r$ ($r>1$), namely, the external input $u$ explicitly appears until $h$ is differentiated $r$ times\footnote{For the simplicity of illustration, we assume that all entries of $u=[u_1,\ldots,u_m]^T$ appear after $h$ is differentiated $r$ times.}.

%
%
%
%
%
%

\begin{definition}[ISSf]\label{def:issf}
System (\ref{eq:system-with-input}) is ISSf on a given set $\setS_0$, if for any $u\in L_\infty^m$ and any $x_0$ in a subset $\X\subseteq\C_0$,  $x(t,x_0,u)$ stays in $\C_0$ for all $t\in I(x_0,u)$.
\end{definition}

\begin{remark}\label{remark:on-ISSf}
The ISSf provides a tool to estimate how the external input $u$ influences the safety. For any $u\in L_\infty^m$, the ISSf of system (\ref{eq:system-with-input}) implies that any $x(t,x_0,u)$ starting from $\setS_0$ may leave this set, but always stays within a finite distance from $\setS_0$ related to the magnitude of $u$ and the ISSf gain $\gamma$. Thus, from the control aspect, an additional safety margin $\gamma(\|u\|)$ should be added to the safety-critical controller so as to avoid the violation of safety constraints. On the other hand, $x(t,x_0,u)$ always stays inside $\setS_0$ if there is no input (i.e., $u\equiv0$).
\end{remark}

\begin{remark}
In contrast to the relative-degree-one results (e.g., \cite{ames2016control, xu2015robustness, kolathaya2018input, lyu2022small}), the trajectory of $h(x(t))$ is not only dependent on the initial value of itself but also the initial value of its high-order derivatives, and thus, $x(t)$ is required to start in a subset of $\C_0$. This assumption has been also employed by the high-relative-degree results \cite{nguyen2016exponential, xu2018constrained, tan2021high, xiao2021high}.
\end{remark}

We then review the set input-to-state stability (set-ISS) that can be used to characterize the robustness of safety when the external input is involved.

\begin{definition}[Set-ISS]\label{def:set-ISS}
System (\ref{eq:system-with-input}) is ISS with respect to a closed set $\setS$, if for any $x_0\in\R^n$ and any $u\in L_\infty^m$, it is forward complete and
\begin{flalign}\label{eq:beta-gamma-ISS}
    |x(t)|_{\setS}\leq\beta(|x_0|_{\setS},t)+\gamma(\|u\|),\;\;\forall t\geq0
\end{flalign}
where $\beta$ is of class $KL$ and $\gamma$ is of class $K$.
\end{definition}

\begin{remark}\label{remark:on-iss}
The set-ISS implies that every state trajectory $x(t)$ always stays within a distance $\beta(|x_0|_{\setS},0)+\gamma(\|u\|)$ from the set $\setS$, and eventually enters within a distance $\gamma(\|u\|)$. In particular, whenever $u\equiv0$, the set-ISS reduces to the set asymptotical stability, and according to \cite[Section 2.2]{xu2015robustness}, any $x(t)$ starting outside $\setS$ will get to this set eventually.
\end{remark}

\begin{remark}\label{remark:on-def-iss}
Note that Definition \ref{def:set-ISS}, different from the set-ISS definitions of \cite{Sontag1995SetISS, shi2011connectivity}, does not require the set $\setS$ to be compact but assumes that system (\ref{eq:system-with-input}) is forward complete. This assumption is reasonable; for example, in the QP-based safety-critical control framework \cite{ames2014control, ames2016control}, boundness of the solution inside and outside the set $\setS$ can be ensured by the control Lyapunov functions (CLFs) and the control barrier functions (CBFs), respectively. Clearly, such an assumption is redundant if $\setS$ is compact.
\end{remark}

\subsection{ISSf-Barrier Functions with High Relative Degree}

For any $C^r$ function $h:\R^n\rightarrow\R$, define
\begin{flalign}\label{eq:eta-k}
    \eta_0(x)=h(x),\;\;\eta_k(x)=\dot{\eta}_{k-1}(x)+\alpha_k(\eta_{k-1}(x)),\;\;1\leq k\leq r
\end{flalign}
where $\alpha_k:\R\rightarrow\R$ is a $C^{r-k}$ extended class $K_\infty$ function.

\begin{definition}\label{def:hoissfbf}
A $C^r$ function $h:R^n\rightarrow\R$ is an ISSf-BF\footnote{In this work, we concentrate on global ISSf-BFs, namely, given a set $\setS_{k-1}=\{x:\eta_{k-1}(x)\geq0\}$, $\eta_{k-1}(x)\rightarrow+\infty$ as $|x|_{\R^n\backslash\setS_{k-1}}\rightarrow+\infty$, and $\eta_{k-1}(x)\rightarrow-\infty$ as $|x|_{\setS_{k-1}}\rightarrow+\infty$ for $k=1,\ldots,r$.} with relative degree $r$ for system (\ref{eq:system-with-input}), if there exists a class $K_\infty$ function $\gamma$ such that (\ref{eq:eta-k}) and
\begin{flalign}\label{eq:eta-r}
    \eta_r(x)\geq-\gamma(|u|)
\end{flalign}
hold for all $x\in\R^n$ and $u\in L_\infty^m$.
\end{definition}

The ISSf-BF in Definition \ref{def:hoissfbf} is a variant of the ZBF of \cite{xu2015robustness} with the consideration of external inputs, and thus, inherits a good property of ZBF that $x(t)$ is allowed to get close to the unsafe region when it is far away from this region. It reduces to the high-order ZBF of \cite[Def. 2]{tan2021high} if $u\equiv0$. Analogous to ISS-Lyapunov functions that have different equivalent definitions, one can redefine the ISSf-BF by replacing (\ref{eq:eta-r}) with
\begin{flalign}
    |\eta_{r-1}(x)|\geq\phi(|u|)\Rightarrow\eta_{r}(x)\geq0.
      \label{eq:margin-barrier}
\end{flalign}
where $\phi$ is a class $K_\infty$ function.

\begin{lemma}\label{lem:equivalence-ISSf-BF}
Inequalities (\ref{eq:eta-r}) and (\ref{eq:margin-barrier}) are equivalent.
\end{lemma}

\noindent
\textbf{Proof.}
See Appendix I.
\hfill $\Box$
\vskip5pt

The analysis in this paper is based on (\ref{eq:eta-k}) and (\ref{eq:eta-r}), while (\ref{eq:margin-barrier}) is also useful, e.g., constructing an inverse optimal safety-critical controller as in \cite{krstic2021inverse}.

Consider the coordinate transformation
\begin{flalign}\label{eq:coordination-transform}
    \tilde\eta_{k-1}=\eta_{k-1}+\hat\alpha_k\circ\gamma(\|u\|),\;\;k=1,\ldots, r
\end{flalign}
where
\begin{flalign}
    \hat\alpha_k(s)=-\alpha^{-1}_{k}\circ\alpha^{-1}_{k+1}\circ\cdots\circ\alpha^{-1}_{r}(-s).
     \nonumber
\end{flalign}
From (\ref{eq:eta-k}) and (\ref{eq:eta-r}), we have the following auxiliary ZBF condition with relative degree $r$:
\begin{flalign}
    &\dot{\tilde\eta}_{k-1}(x)=-\mu_k(\tilde\eta_{k-1}(x))+\tilde\eta_{k}(x),\;\;k=1,\ldots,r-1
     \label{eq:axui-barrier-cond-A}\\
    &\dot{\tilde\eta}_{r-1}(x)\geq-\mu_r(\tilde\eta_{r-1}(x))
     \label{eq:axui-barrier-cond-B}
\end{flalign}
where $\mu_k(s):=\alpha_k(s-\hat\alpha_{k}\circ\gamma(\|u\|))+\hat\alpha_{k+1}\circ\gamma(\|u\|)$ and $\mu_r(s):=\alpha_r(s-\hat\alpha_r\circ\gamma(\|u\|))+\gamma(\|u\|)$ are zero at zero and strictly increasing, and thus, are of extended class $K_\infty$. Define the sets
\begin{flalign}
    &\setS_{k-1}=\{x\in\R^n:\eta_{k-1}(x)\geq0\},
     \label{eq:smallset-k-cascade}\\
    &\C_{k-1}=\{x\in\R^n:\tilde\eta_{k-1}(x)\geq0\}.
     \label{eq:largerset-k-cascade}
\end{flalign}
Then we have the main result of this section as follows.

\begin{theorem}\label{thm:HOISSF-BF}
Consider system (\ref{eq:system-with-input}) with safety constraints characterized by $\setS_0$. Suppose $h:\R^n\rightarrow\R$ is an ISSf-BF with relative degree $r$, and satisfies (\ref{eq:eta-k}) and (\ref{eq:eta-r}). Then,
\begin{basedescript}{\desclabelstyle{\pushlabel}\desclabelwidth{0.7cm}}
\item[\hspace{0.17cm}(i)] system (\ref{eq:system-with-input}) is ISSf on $\setS_0$, and the set $\C=\bigcap_{k=1}^{r}C_{k-1}$ is robustly forward invariant;
\item[\hspace{0.17cm}(ii)] system (\ref{eq:system-with-input}) is asymptotically stable with respect to $\C$, and is ISS with respect to the set $\setS=\bigcap_{k=1}^{r}\setS_{k-1}$ if it is forward complete.
\end{basedescript}
\end{theorem}

\noindent
\textbf{Proof.} See Appendix II.
\hfill $\Box$
\vskip5pt

\begin{remark}
The proof of Theorem \ref{thm:HOISSF-BF} is challenging compared with the relative-degree-one result in \cite[Theorem 1]{kolathaya2018input} because of the high-order derivatives involved in the ISSf-BFs, as can be seen in (\ref{eq:eta-k}) and (\ref{eq:eta-r}). To handle this issue, we introduce the coordination transform (\ref{eq:coordination-transform}) to establish the relationship between the ISSf-BF in Definition \ref{def:hoissfbf} and the high-order ZBF of \cite{tan2021high}. In this way, we can prove Theorem \ref{thm:HOISSF-BF} by analyzing auxiliary ZBF condition (\ref{eq:axui-barrier-cond-A}) and (\ref{eq:axui-barrier-cond-B}) instead of the original ISSf-BF. This analysis simplifies the proof and provides new insight for ISSf verification under high-relative-degree safety constraints. On the other hand, the region $\C\backslash\setS$ is smaller for larger $\alpha_1$, \ldots, $\alpha_r$. Thus, one can select large $\alpha_1$, \ldots, $\alpha_r$ to improve the robustness of safety against the uncertainties resulting from the external input $u$. However, as shown in \cite[Section 4.1]{kong2013exponential}, large functions $\alpha_1$, \ldots, $\alpha_r$ will make the computation of barrier functions encounter numerical problems.
\end{remark}

%

\section{Small-Gain Theorem for Safety Verfication}\label{sec:small-gain}

The purpose of this section is to develop a small-gain theorem for the safety analysis and verification of the following interconnected system with high-relative-degree safety constraints:
\begin{flalign}\label{eq:inter-sys-with-input}
    \dot{x}_1=f_1(x_1,x_2,u_1),\;\;\dot{x}_2=f_2(x_1,x_2,u_2),
\end{flalign}
where $x_i\in\R^{n_i}$ and $u_i\in L_\infty^{m_i}$ for $i=1,2$. Let $n=n_1+n_2$, $x=[x_1^T, x_2^T]^T$, $x_0=[x_1(0)^T, x_2(0)^T]^T$ and $u=[u_1^T,u_2^T]^T$.

Given a $C^r$ function $h_i:\R^{n_i}\rightarrow\R$, define
\begin{flalign}\label{eq:eta-nonlinear-sg}
    \eta_{i,0}(x_i)=h_i(x_i),
     \;\;
      \eta_{i,k}(x_i)=\dot{\eta}_{i,k-1}(x_i)+\alpha_{i,k}(\eta_{i,k-1}(x_i))
\end{flalign}
for $i=1,2$ and $k=1$, \ldots, $r$, where $\alpha_{i,k}$ is a $C^{r-k}$ extended class $K_\infty$ function. Suppose that $h_i$ is an ISSf-BF for the $x_i$-system with
\begin{subequations}\label{eq:issf-sg-externalinput}
\begin{flalign}
    &\eta_{1,r}(x_1)\geq\phi_1(h_2(x_2))-\gamma_1(|u_1|),
     \label{eq:issf-sg-externalinput-A}\\
    &\eta_{2,r}(x_2)\geq\phi_2(h_1(x_1))-\gamma_2(|u_2|)
     \label{eq:issf-sg-externalinput-B}
\end{flalign}
\end{subequations}
where $\phi_i$ is of extended class $K_\infty$ and $\gamma_i$ is of class $K_\infty$. Let
\begin{flalign}\label{eq:dik}
    d_{i,k-1}=\min\{\hat\phi_{i,k}(-\hat\gamma_{3-i,1}(\|u\|)),-\hat\gamma_{i,k}(\|u\|)\},
\end{flalign}
where
\begin{flalign}
    &\hat\phi_{i,k}(s)=(\Id+\sigma)\circ\alpha^{-1}_{i,k}\circ\cdots
     \nonumber\\
    &\;\;\;\;\;\;\;\;\;\;\;\;\;\;\;\;\;\;\;\;
    \circ(\Id+\sigma)\circ\alpha^{-1}_{i,r}\circ(\Id+\sigma)\circ\phi_i(s),
     \label{eq:hat-phi-ik}\\
    &\hat\gamma_{i,k}(s)=-(\Id+\sigma)\circ\alpha^{-1}_{i,k}\circ\cdots
     \nonumber\\
    &\;\;\;\;\;\;\;\;\;\;\;\;\;\;\;\;\;\;\;\;
    \circ(\Id+\sigma)\circ\alpha^{-1}_{i,r}\circ(\Id+\sigma^{-1})(-\gamma_i(s))
     \label{eq:gamma-i-k}
\end{flalign}
with $\sigma$ of extended class $K_\infty$. Define the set
\begin{flalign}
    &\setS_{i,k-1}=\{x\in\R^{n}:\eta_{i,k-1}(x_i)\geq0\},
     \label{eq:sg-safe-set}\\
    &\C_{i,k-1}=\{x\in\R^n:\eta_{i,k-1}(x_i)\geq d_{i,k-1}\}.
     \label{eq:sg-larger-set}
\end{flalign}
Because $\hat\phi_{i,k}$ and $\hat\gamma_{i,k}$ are of extended class $K_\infty$ and of class $K_\infty$, respectively, $d_{i,k}\leq0$ for any $u\in L_\infty^{m_1+m_2}$, and thus, $\setS_{i,k-1}\subseteq\C_{i,k-1}$.

\subsection{Comparison Technique}

The following lemma provides a useful comparison technique for establishing the result of this section.

\begin{lemma}\label{lem:alpha-to-beta}
Let $\eta:[0,T)\rightarrow\R$ be a continuous function such that
\begin{flalign}\label{eq:eta-geq-sub-eta-w}
    \dot\eta(t)\geq-\alpha(\eta(t))+w(t),\;\;\forall t\in[0,T)
\end{flalign}
with $\eta(0)=\eta_0$, where $\alpha$ is a locally Lipschitz extended class $K_\infty$ function, and $w:\R_{\geq0}\rightarrow\R$ is a locally essentially bounded function. Then there exists an extended class $KL$ function $\beta:\R\times\R_{\geq0}\rightarrow\R$ with $\beta(s,0)=s$ such that
\begin{flalign}\label{eq:compar-lemma-eta-low-bd}
    \eta(t)
    \geq\beta(\eta_0-\eta^*,t)
    +\eta^*,
    \;\;\forall t\in[0,T)
\end{flalign}
where $\eta^*=\alpha^{-1}(\inf_{t\in[0,T)}w(t))$.
\end{lemma}
\noindent
\textbf{Proof.} See Appendix III.
\hfill $\Box$
\vskip5pt

A direct application of Lemma \ref{lem:alpha-to-beta} is to prove \cite[Theorem 1]{kolathaya2018input}. To see this, we consider a system with solutions defined on $[0,T)$ and a relative-degree-one ISSf-BF $h:\R^n\rightarrow\R$ satisfying
\begin{flalign}
    \dot h(x)=-\alpha(h(x))-\gamma(|u|)
\end{flalign}
where $\alpha$ is of extended class $K_\infty$, and $\gamma$ is of class $K_\infty$. Because $\gamma(|u|)\leq\gamma(\|u\|)$, it follows from Lemma \ref{lem:alpha-to-beta} (by taking $w(t)=-\gamma(|u(t)|)$) that
\begin{flalign}\label{eq:low-bd-degreeone-issf-bf}
    h(x(t))
    &\geq\beta(h(x_0)-\alpha^{-1}(-\gamma(\|u\|)),t)
     \nonumber\\
    &\;\;\;\;\;\;\;\;\;\;\;\;\;\;\;\;\;\;\;\;\;\;\;\;
     +\alpha^{-1}(-\gamma(\|u\|)),\;\;\forall t\in[0,T)
\end{flalign}
where $\beta$ is of extended class $KL$. Thus, $x(t)$ always stays inside the set $\C=\{x:h(x)-\alpha^{-1}(-\gamma(\|u\|))\geq0\}$ if $x_0\in\C$.

\begin{remark}
As can be seen in (\ref{eq:low-bd-degreeone-issf-bf}), Lemma \ref{lem:alpha-to-beta} provides an estimate on how the lower bound of $h$ is influenced by the external input $u$. Also, it provides an ISSf analysis approach from an input-output viewpoint if we treat the ISSf-BF $h$ as an output function. An advantage of this technique is that it does not require any forward completeness assumptions, which is particularly useful for the ISSf analysis of interconnected systems because it is easy for an interconnected system to have a finite escape time. On the other hand, as shown in (\ref{eq:low-bd-degreeone-issf-bf}), the first argument of $\beta$ contains the initial condition and the boundary of $\C$, which helps us explicitly analyze the influence of the initial condition and the boundary of $\C$ on safety. It is interesting to note that this estimate is less conservative than that of \cite[Definition 2]{krstic2021inverse}, where the lower bound of $h(x)$ is estimated as
\begin{flalign}\label{eq:low-bd-degreeone-issf-bf-krstic}
    h(x(t))
    \geq\beta(h(x_0),t)
     +\alpha^{-1}(-\gamma(\|u\|)),\;\;\forall t\in[0,T).
\end{flalign}
To see this, we select an initial condition $x_0$ such that $\alpha^{-1}(-\gamma(\|u\|))\leq h(x_0)<0$. From (\ref{eq:low-bd-degreeone-issf-bf-krstic}), $x(t)$ with such an initial condition may leave the set $\C$, which is actually not the case according to \cite[Theorem 1]{kolathaya2018input}.
\end{remark}

\subsection{Small-Gain Theorem under High-Relative-Degree Safety Constraints}

The following result provides a small-gain theorem to ensure that the interconnection of two ISSf systems is still ISSf under high-relative-degree constraints.

\begin{theorem}\label{thm:non-sg-with-input}
Consider the interconnected system (\ref{eq:inter-sys-with-input}) with safety constraints characterized by $\setS_{1,0}\bigcap\setS_{2,0}$. Let $J(x_0,u)$ be the maximal interval on which the distance between $x(t)$ and  the unsafe region $\R^n\backslash(\setS_{1,0}\bigcap\setS_{2,0})$ is finite. Suppose that, for $i=1,2$, the $x_i$-system has an ISSf-BF $h_i$ satisfying (\ref{eq:eta-nonlinear-sg}) and (\ref{eq:issf-sg-externalinput}). If
\begin{flalign}\label{eq:sg-cond}
    |\hat\phi_{1,1}\circ\hat\phi_{2,1}(s)|<|s|,\;\;\forall s\in\R\backslash\{0\},
\end{flalign}
then
\begin{basedescript}{\desclabelstyle{\pushlabel}\desclabelwidth{0.7cm}}
\item[\hspace{0.17cm}(i)] the solution $x(t)$ is right maximally defined on $I(x_0,u)=J(x_0,u)$;
\item[\hspace{0.17cm}(ii)]  system (\ref{eq:inter-sys-with-input}) is ISSf on $\setS_{1,0}\bigcap\setS_{2,0}$, the set $\C=\bigcap_{i=1,2}\bigcap_{k=1}^{r}\C_{i,k-1}$ is robustly forward invariant;
\item[\hspace{0.17cm}(iii)] system (\ref{eq:inter-sys-with-input}) is ISS with respect to $\setS=\bigcap_{i=1,2}\bigcap_{k=1}^{r}\setS_{i,k-1}$ if $J(x_0,u)=\R_{\geq0}$.
\end{basedescript}
\end{theorem}

\noindent
\textbf{Proof.} See Appendix IV.
\hfill $\Box$
\vskip5pt

The following remarks discuss the assumptions, conclusions, contributions and challenges of Theorem \ref{thm:non-sg-with-input}.

\begin{remark}[Reasonableness of the Assumption]
The assumption on the finite distance from $x(t)$ to the unsafe region $\R^n\backslash(\setS_{1,0}\bigcap\setS_{2,0})$ implies that system (\ref{eq:inter-sys-with-input}) does not have a finite escape time whenever $x(t)$ is inside the safe set. As indicated in Remark \ref{remark:on-def-iss}, this assumption can be guaranteed by CLFs in the well-known QP-based safety-critical control framework \cite{ames2014control, ames2016control}.
\end{remark}

\begin{remark}[Comparison with Existing Results]
There are two differences between Theorem \ref{thm:non-sg-with-input} and the results of \cite{lyu2022small,jagtap2020compositional}. Firstly, Theorem \ref{thm:non-sg-with-input} allows the safety constraints have high relative degree, and thus, can be used to handle the complicated safety-critical control problems (see, e.g., the pendulum-spring-cart system with output constraints given in Section \ref{sec:example}). Secondly, we further verify the ISS of interconnected system (\ref{eq:inter-sys-with-input}) with respect to $\setS$ (a subset of the compositional safe set $\setS_{1,0}\bigcap\setS_{2,0}$). Thus, whenever there is no external input, any $x(t)$ staring outside $\setS_{1,0}\bigcap\setS_{2,0}$ will become safe eventually, as discussed in Remark \ref{remark:on-iss}.
\end{remark}

\begin{remark}[Challenges of the Proof]
Compared with the relative-degree-one results \cite{lyu2022small,jagtap2020compositional}, the proof of Theorem \ref{thm:non-sg-with-input} is more challenging. In \cite{lyu2022small}, the safety of interconnected systems is verified by analyzing the derivatives of individual ISSf-BFs on the boundary of the compositional safe set. In \cite{jagtap2020compositional}, a discrete-time compositional ISSf-BF is constructed to verify safety with the help of the small-gain condition and the assumption that the state trajectories of subsystems cannot get close to the boundary of safe set. However, because the individual ISSf-BFs of Theorem \ref{thm:non-sg-with-input} contain a set of high-order derivatives and the state trajectories are allowed to get close to the unsafe regions, it is difficult to analyze the derivatives of individual ISSf-BFs or construct a compositional ISSf-BF.
\end{remark}

\begin{figure}[!htb]
 \centering
 \includegraphics[width=8.8cm]{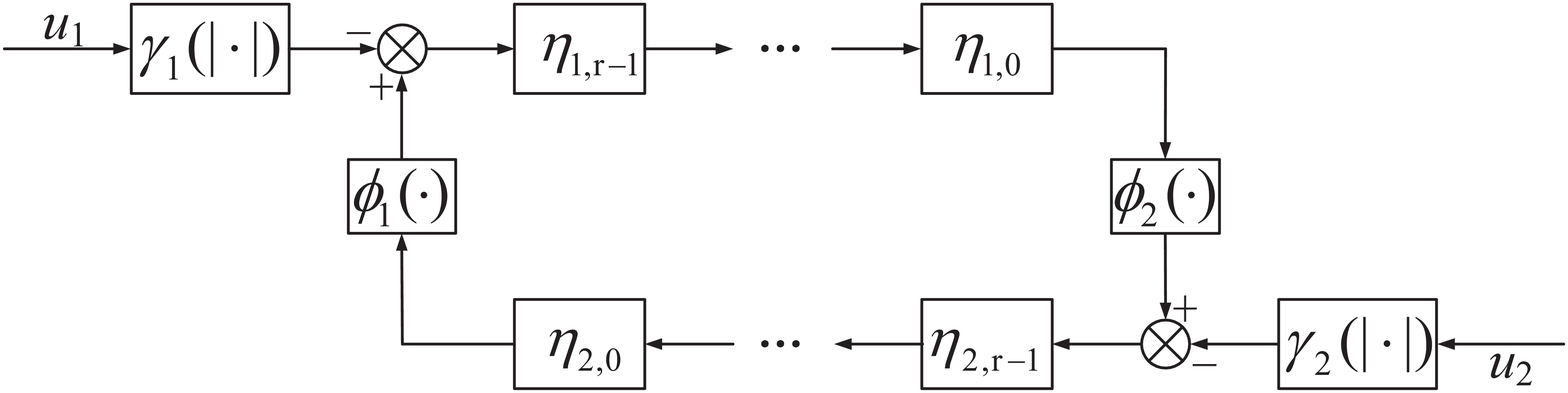}
 \caption{Visual illustration of barrier condition (\ref{eq:eta-nonlinear-sg}) and (\ref{eq:issf-sg-externalinput}).}
\label{Fig:ISSf-BF}
\end{figure}

\begin{remark}[Main Ideas for Proving Theorem \ref{thm:non-sg-with-input}]
As indicated in Fig. \ref{Fig:ISSf-BF}, the barrier condition (\ref{eq:eta-nonlinear-sg}) and (\ref{eq:issf-sg-externalinput}) is a feedback loop consisting of two chains interconnected with each other. For each chain, $(x_{3-i},u_i)$ is the input, $h_i(x_i)=\eta_{i,0}(x_i)$ is the output, and the ``$\eta_{i,k-1}$-systems'' (containing the high-order derivatives of $h_i$) are cascaded with each other. In fact, the analysis of safety and set ISS is essentially equivalent to analyzing the lower bound and the convergence of $h_i(x_i(t))$, as detailed in Appendix IV. This observation motivates us to prove Theorem \ref{thm:non-sg-with-input} from an input-output viewpoint, instead of analyzing the derivatives of ISSf-BFs or constructing a compositional ISSf-BF. Specifically, the proof is divided into the following three steps.
\begin{itemize}
  \item \textbf{Step 1:} Treat $\eta_{i,k}$ and $\eta_{i,k-1}$ as the input and output of the ``system'' $\dot{\eta}_{i,k-1}=-\alpha_{i,k}(\eta_{i,k-1})+\eta_{i,k}$, and apply Lemma \ref{lem:alpha-to-beta} to estimate how the lower bound and the convergence of $\eta_{i,k-1}$ are influenced by $\eta_{i,k}$.
  \item \textbf{Step 2:} For each chain, establish the relationship between its input $(x_{3-i},u_i)$ and the lower bound or the convergence of its output $h_i(x_i)$ recursively with the lower bounds of $\eta_{i,0}$, \ldots, $\eta_{i,r-1}$ estimated in Step 1.
  \item \textbf{Step 3:} Use the small-gain condition (\ref{eq:sg-cond}) to cancel the influence of the feedback interconnection so as to make that the lower bound and the convergence of $h_i(x_i)$ are only dependent on the input $(u_1,u_2)$ of interconnected system (\ref{eq:inter-sys-with-input}).
\end{itemize}

\end{remark}

Note that, for the case $\phi_1(s)=0$ or $\phi_2(s)=0$, the small-gain condition (\ref{eq:sg-cond}) always holds. Thus, we have the following corollary for the cascade connection of two ISSf subsystems.

\begin{corollary}
Consider the cascade system
\begin{flalign}\label{eq:cascade-sys-with-input}
    \dot{x}_1=f_1(x_1,x_2,u_1),\;\;\dot{x}_2=f_2(x_2,u_2)
\end{flalign}
with safety constraints characterized by $\setS_{1,0}\bigcap\setS_{2,0}$. Let $J(x_0,u)$ be the maximal interval on which the distance between $x(t)$ and  the unsafe region $\R^n\backslash(\setS_{1,0}\bigcap\setS_{2,0})$ is finite.
Suppose that $h_1$ and $h_2$ are ISSf-BFs satisfying (\ref{eq:eta-nonlinear-sg}), (\ref{eq:issf-sg-externalinput-A}) and
\begin{flalign}
    \eta_{2,r}(x_2)\geq-\gamma_2(|u_2|).
    \nonumber
\end{flalign}
Then the conclusions of Theorem \ref{thm:non-sg-with-input} also holds for system (\ref{eq:cascade-sys-with-input}) with $d_{2,k-1}$ in (\ref{eq:dik}) modified as $d_{2,k-1}=-\hat\gamma_{2,k}(\|u\|)$.
\end{corollary}

\section{Illustrative Example}\label{sec:example}

In this section, the effectiveness of the proposed small-gain technique is illustrated on the decentralized tracking control of the pendulum-spring-cart system \cite[Sec. 7]{shi1992decentralized}:
\begin{subequations}\label{eq:example-mechan}
\begin{flalign}
    &\dot x_{i,1}=x_{i,2}
     \\
    &\dot x_{i,2}=\frac{g}{wl}x_{i,1}
                    -\frac{m}{M}x_{i,2}^2\sin x_{i,1}
                    -\frac{a(t)k(a(t)-wl)}{wml^2}x_{i,1}
     \nonumber\\
    &\;\;\;\;\;\;
                    +\frac{kb(a(t)-wl)}{wml^2}
                    +\frac{1}{wml^2}u_i
                    +\frac{a(t)k(a(t)-wl)}{wml^2}x_{3-i,1}
\end{flalign}
\end{subequations}
for $i=1,2$, where $x_{i,1}=\theta_i$ and $x_{i,2}=\dot{\theta}_i$ denote the angular displacement and the angular velocity, respectively, $u_i$ is the control torque applied to the pendulum, $m$ and $l$ are the mass and the length of the pendulum, $M$ is the mass of the car, $w = m/(M + m)$, $k$ is the spring constant, $L$ is natural length of the spring, $a(t)\in[0,l]$ is the distance from the pivot of the spring to the bottom of the pendulum, $g$ is the gravitational acceleration, and $b$ is the distance between the cars. Choose $g$ = 9.8 $\text{m}/\text{s}^2$, $l$ = 1 m, $k$ = 1 n/m. $M$ = 15 kg, $m$ = 5 kg, $b$ = 2 m and $a$ = $0.75$ m.

Suppose that the safety constraint of pendulum $i$ is $\theta_i(t)\geq\underline{\theta}_i$, where $\underline{\theta}_i\geq0$ denotes the lower bound of $\theta_i(t)$. The goal is to make the output $\theta_i(t)$ of the pendulum track its own reference trajectory $y_{r,i}$, while simultaneously avoiding the violation of safety constraints.

\subsection{Nominal Tracking Controller}

We design a nominal tracking controller with the backstepping technique \cite{krstic1995nonlinear}. Consider the coordination transform
\begin{flalign}
    z_{i,1}=x_{i,1}-y_{r,i},\;\;z_{i,2}=x_{i,2}-\varpi_i
    \nonumber
\end{flalign}
where $\varpi_i=-r_{i,1}z_{i,1}+\dot y_{r_i}$ with $r_{i,1}>0$ as a designed parameter. Then the nominal controller is chosen as
\begin{flalign}\label{eq:nominal-control}
    \hat u_i=wml^2\Big(-r_{i,2}z_{i,2}-W_i-\frac{ak(a-wl)}{2wml^2}(z_{i,2}+2y_{r,3-i})\Big)
\end{flalign}
where $r_{i,2}>0$ is a designed parameter, and
\begin{flalign}
    W_i
    &=z_{i,1}
     -\dot\varpi_{i}
     +\frac{g}{wl}x_{i,1}
     -\frac{m}{M}x_{i,2}^2\sin x_{i,1}
     \nonumber\\
    &\;\;\;\;\;\;\;\;\;\;\;\;\;\;\;\;
     -\frac{ak(a-wl)}{wml^2}x_{i,1}
     +\frac{kb(a-wl)}{wml^2}.
     \nonumber
\end{flalign}
We can verify that the derivative of the Lyapunov function candidate $V_i=(z_{i,1}^2+z_{i,2}^2)/2$ along the solution of the closed-loop system consisting of (\ref{eq:example-mechan}) and (\ref{eq:nominal-control}) satisfies
\begin{flalign}
    \dot V_i\leq-\lambda_i V_i+\chi_i(V_{3-i})
    \nonumber
\end{flalign}
where $\lambda_i:=\min\{r_{i,1},r_{i,2}\}$ and $\chi_i(s):=\frac{ak(a-wl)}{2wml^2}s$. Choose sufficiently large $\lambda_i$ for $i=1,2$, such that $\chi_1(\chi_2(s)/\lambda_2)/\lambda_1<s$ for all $s>0$, and according to \cite[Theorem 5.1]{jiang1999combined}, the tracking error $z_{i,1}$ is driven to zero.

\subsection{Control Barrier Function}

Let $h_i(x_i)=x_{i,1}-\underline{\theta}_i$, which is clearly with relative degree two. Then we can establish (\ref{eq:eta-nonlinear-sg}) with $\eta_{i,1}=x_{i,2}+\alpha_{i,1}(h_i)$ and
\begin{flalign}\label{eq:eta-i2-example}
    \eta_{i,2}
    &=\dfrac{g}{wl}x_{i,1}
                -\dfrac{m}{M}x_{i,2}^2\sin x_{i,1}
                +\alpha_{i,1}'(h_{i})x_{i,2}
                +\alpha_{i,2}(\eta_{i,1})
     \nonumber\\
    &\;\;\;\;
                -\dfrac{k(a-wl)}{wml^2}(ax_{i,1}-b-\underline\theta_{3-i})
                +\dfrac{1}{wml^2}u_i
                +\phi_i(h_{3-i})
\end{flalign}
where $\phi_i(s):=\dfrac{ak(a-wl)}{wml^2}s$, and $\alpha_{i,1}$ and $\alpha_{i,2}$ are used to tune the ISSf gain so as to satisfy the small-gain condition (\ref{eq:sg-cond}). Let $\psi_{i,1}(x_i)=\dfrac{g}{wl}x_{i,1}
                                                -\dfrac{m}{M}x_{i,2}^2\sin x_{i,1}
                                                +\alpha_{i,1}'(h_{i})x_{i,2}
                                                +\alpha_{i,2}(\eta_{i,1})
                                                -\dfrac{k(a-wl)}{wml^2}(ax_{i,1}-b-\underline\theta_{3-i})$
and $\psi_{i,0}(x_i)=\dfrac{1}{wml^2}$. Then (\ref{eq:eta-i2-example}) can be rewritten as
\begin{flalign}
    \eta_{i,2}=\psi_{i,1}(x_i)+\psi_{i,0}(x_i)u_i+\phi_i(h_{3-i}(x_{3-i})).
    \nonumber
\end{flalign}
Inspired by the control barrier function \cite{ames2016control, xu2015robustness, ames2014control, wieland2007constructive}, any control input $u_i$ in the set
\begin{flalign}\label{eq:issf-cbf}
    U_i = \{u_i\in\R:\psi_{i,1}(x_i)+\psi_{i,0}(x_i)u_i\geq0\}
\end{flalign}
renders
\begin{flalign}
    \eta_{i,2}(x_i)\geq\phi_i(h_{3-i}(x_{3-i})).
    \nonumber
\end{flalign}
Take $\alpha_{i,k}(s)=c_{i,k}s$ for $i=1,2$ and $k=1,2$. Select sufficiently larger $c_{i,k}$ such that (\ref{eq:sg-cond}) is satisfied. Because no external input is involved in the closed-loop system (\ref{eq:example-mechan}), it follows from Theorem \ref{thm:non-sg-with-input} that i) if $\theta_1(0)\geq0$ and $\theta_{2}(0)\geq0$, then the angular displacements $\theta_1(t)$ and $\theta_2(t)$ do not violate the safety constraints; and ii) if $\theta_1(0)<0$ or $\theta_{2}(0)<0$, then the closed-loop system will be safe eventually.

\subsection{Simulation Results}

\begin{figure}[!htb]
 \centering
 \includegraphics[width=8cm]{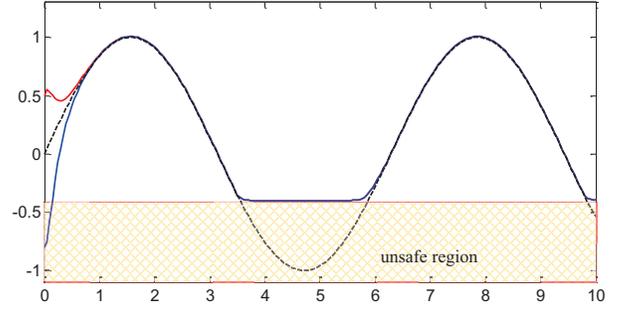}
 \caption{Tracking result of $\theta_1(t)$.}
\label{Fig:traj-1}
\end{figure}

\begin{figure}[!htb]
 \centering
 \includegraphics[width=8cm]{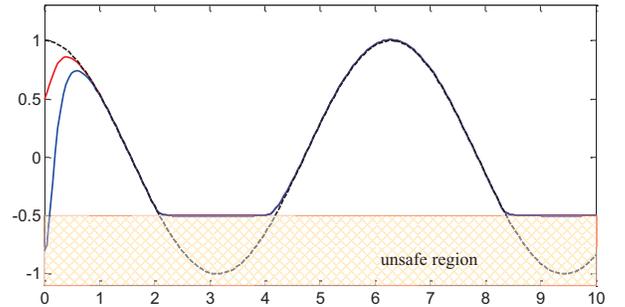}
 \caption{Tracking result of $\theta_2(t)$.}
\label{Fig:traj-2}
\end{figure}

According to (\ref{eq:nominal-control}) and (\ref{eq:issf-cbf}), we can establish the QP-based controller as in \cite{ames2016control, ames2014control}:
\begin{flalign}
    u_i^*=
    &\mathop{\arg\min}_{u\in\R}|u_i-\hat u_i|,\;\;\;\;
     \nonumber\\
    &\text{s.t.}\;\; \psi_{i,1}(x_i)+\psi_{i,0}(x_i)u_i\geq0.
    \nonumber
\end{flalign}
Set $\underline{\theta}_1=-0.4$, $\underline{\theta}_2=-0.5$, $y_{r,1}=\sin(t)$ and $y_{r,2}=\cos(t)$. Choose the design parameters as: $r_{1,1}=r_{2,1}=10$, $r_{1,2}=r_{2,2}=5$, $c_{1,1}=c_{2,1}=20$, and $c_{1,2}=c_{2,2}=10$. The simulation results are given in Figs. \ref{Fig:traj-1} and \ref{Fig:traj-2}, where the black dash line denotes the reference trajectory, the red and the blue solid lines represent the tracking results of $\theta_i(t)$ with initial conditions $(x_{i,1}(0),x_{i,2}(0))=(0.5, 1.0)$ and $(x_{i,1}(0),x_{i,2}(0))=(-0.8, 1.0)$, respectively. Clearly, the tracking task is achieved if the reference signal is inside the safe region. Moreover, for the simulation with $(x_{i,1}(0),x_{i,2}(0))=(0.5, 1.0)$, $\theta_i(t)$ always stays inside the safe region, while, for the other one, $\theta_i(t)$ enters the safe region eventually without violating the safety constraint any more, even though it is initialized unsafely.

\section{Conclusions}\label{sec:concl}

In this work, we developed a small-gain technique for the safety verification of interconnected systems under high-relative-degree safety constraints. The ISSf-BFs was used to capture the safety of subsystems. With the help of high-relative-degree ISSf-BFs, a small-gain theorem was given for the safety analysis and verification of interconnected systems. Finally, the decentralized control of a pendulum-spring-cart system with output constraints was used to illustrate the effectiveness of our result.

\section*{Appendix I: Proof of Lemma \ref{lem:equivalence-ISSf-BF}}

\setcounter{subsection}{0}

(\ref{eq:eta-r}) $\Rightarrow$ (\ref{eq:margin-barrier}). According to (\ref{eq:eta-k}) and (\ref{eq:eta-r}), we have
\begin{flalign}
    &\eta_{r-1}(x)\leq\alpha_r^{-1}(-\gamma(|u|)/c)\Rightarrow\dot\eta_{r-1}(x)+(1-c)\alpha_r(\eta_{r-1}(x))\geq0,
     \nonumber\\
    &\eta_{r-1}(x)\geq\alpha_r^{-1}(\gamma(|u|)/c)\Rightarrow\dot\eta_{r-1}(x)+(1+c)\alpha_r(\eta_{r-1}(x))\geq0
     \nonumber
\end{flalign}
where $c\in(0,1)$ is a constant. Let
\begin{flalign}
    &\phi(s)=\max\{-\alpha_r^{-1}(-\psi(s)/c),\alpha_r^{-1}(\psi(s)/c)\},
     \nonumber\\
    &\hat\alpha_r(s)=\max\{(1-c)\alpha_r(s),(1+c)\alpha_r(s)\}.
     \nonumber
\end{flalign}
Clearly, $\phi$ is a class $K_\infty$ function on $\R_{\geq0}$, and $\hat\alpha$ is a $C^0$ extended class $K_\infty$ function on $\R$. Thus, (\ref{eq:margin-barrier}) follows by taking $\eta_r(x)=\dot\eta_{r-1}(x)+\hat\alpha_r(\eta_{r-1}(x))$.

(\ref{eq:margin-barrier}) $\Rightarrow$ (\ref{eq:eta-r}). According to (\ref{eq:margin-barrier}), if $|\eta_{r-1}(x)|\leq\phi(|u|)$, then
\begin{flalign}
    \eta_{r}(x)
    &=\alpha_r(\eta_{r-1}(x))+\nabla\eta_{r-1}(x)f(x,u)
    \nonumber\\
    &\geq\alpha_r(-\phi(|u|))+\inf_{|\eta_{r-1}(x)|\leq\phi(|u|)}\nabla\eta_{r-1}(x)f(x,u)
     \nonumber\\
    &\geq-\gamma(|u|)
     \nonumber
\end{flalign}
where
\begin{flalign}
    \gamma(s)=-\alpha_r(-\phi(s))
    -\inf_{|\eta_{r-1}(x)|\leq\phi(s)}\min\{0,\nabla\eta_{r-1}(x)f(x,s)\}.
    \nonumber
\end{flalign}
Because $\phi$ is of extended class $K_\infty$, the set $\{x\in\R^n:|\eta_{r-1}(x)|\leq\phi(s)\}$ is compact for fixed $r\geq0$, and thus, $\inf_{|\eta_{r-1}(x)|\leq\phi(s)}\min\{0,\nabla\eta_{r-1}(x)f(x,s)\}$ is well defined, non-negative and non-increasing for all $s\geq0$. On the other hand, if $|\eta_{r-1}(x)|\geq\phi(|u|)$, $\eta_{r}(x)\geq0\geq-\gamma(|u|)$. According to (\ref{eq:eta-k}) and (\ref{eq:margin-barrier}), $\nabla\eta_{r-1}(x)f(x,u)\geq-\alpha(\eta_{r-1}(x))\geq0$ whenever $\eta_{r-1}(x)=0$ and $u=0$. Thus, $\inf_{|\eta_{r-1}(x)|\leq\phi(s)}\min\{0,\nabla\eta_{r-1}(x)f(x,s)\}$ is zero at $s=0$, and consequently, $\gamma$ is of class $K_\infty$.

\section*{Appendix II: Proof of Theorem \ref{thm:HOISSF-BF}}

\setcounter{subsection}{0}

\subsection{Proof of (i) of Theorem \ref{thm:HOISSF-BF}}

By applying Proposition 1 of \cite{tan2021high} to the auxiliary ZBF condition (\ref{eq:axui-barrier-cond-A}) and (\ref{eq:axui-barrier-cond-B}), we have
\begin{flalign}
    \tilde\eta_{k-1}(x(t))=\eta_{k-1}(x(t))+\hat\alpha_k\circ\gamma(\|u\|)\geq0,\;\forall x_0\in\C
    \nonumber
\end{flalign}
for $k=1,\ldots,r$, and thus, the set $\C$ is robustly forward invariant. Because $\C$ is a subset of $\C_0$, $x(t,x_0,u)$ cannot leave $\C_{0}$ for any $x_0\in\C$, which further implies the ISSf of system (\ref{eq:system-with-input}) on the set $\setS_0$.

\subsection{Proof of (ii) of Theorem \ref{thm:HOISSF-BF}}

Let
\begin{flalign}\label{eq:single-lya}
    \tilde{V}_{k-1}(x)=\max\{0,-\tilde\eta_{k-1}(x)\},\;\;k=1\ldots,r.
\end{flalign}
Since $-\eta_{k-1}(x)\leq0$ whenever $x\in\C_{k-1}$, (\ref{eq:single-lya}) is equivalent to
\begin{flalign}
    V_{k-1}(x)=
    \left\{
      \begin{array}{ll}
        0, & \text{if } x\in \C_{k-1}; \\
        -\eta_{i,k-1}(x_i), & \text{if } x\in\R^{n}\backslash \C_{k-1}.
      \end{array}
    \right.
    \nonumber
\end{flalign}
Because $\tilde{V}_{k-1}\geq-\tilde\eta_{k-1}$, it follows from (\ref{eq:axui-barrier-cond-A}) and (\ref{eq:axui-barrier-cond-B}) that
\begin{flalign}
    &\dot{\tilde V}_{k-1}(x)\leq\mu_k(-\tilde V_{k-1}(x))+\tilde V_{k}(x),\;\;k=1,\ldots,r-1,
     \label{eq:axui-lya-cond-A}\\
    &\dot{\tilde V}_{r-1}(x)\leq\mu_r(-\tilde V_{r-1}(x)).
     \label{eq:axui-lya-cond-B}
\end{flalign}
Consider the comparison system
\begin{flalign}\label{compar-sys-A}
    \left[
      \begin{array}{c}
        \dot m_0 \\
        \dot m_1 \\
        \cdots \\
        \dot m_{r-1} \\
      \end{array}
    \right]
    =
    \left[
      \begin{array}{c}
        \mu_1(-m_0)+m_1 \\
        \mu_2(-m_1)+m_2 \\
        \cdots \\
        \mu_r(-m_{r-1}) \\
      \end{array}
    \right]
\end{flalign}
with $[m_0(0),\ldots,m_{r-1}(0)]^T=[\tilde V_0(x(0)),\ldots,\tilde V_{r-1}(x(0))]^T$. For notational convenience, we take $m=[m_0,\ldots,m_{r-1}]^T$. Because the vector field $F$ is quasi-monotone increasing\footnote{As indicated in \cite[p.314]{rouche1977stability}, a vector field $F:\R^r\rightarrow\R^r$ is said to be quasi-monotone increasing, if $F_k(x)\geq F_k(y)$ for every $k=1$, \ldots, $r$ and any two points $x,y\in\R^r$ satisfying i) $x_p=y_p$ if $p=k$, and ii) $x_p\geq y_{p}$ if $p\neq k$. Herein, the subscript represents the index of entries.}, by the vectorial comparison lemma (see, e.g., Lemma 2.3 of \cite[Chapter 9]{rouche1977stability}), $\tilde V_{k-1}(x(t))\leq m_{k-1}(t)$ for all $t\geq0$ with $k=1,\ldots,r$. Moreover, from Proposition 3 of \cite{tan2021high}, system (\ref{compar-sys-A}) is asymptotically stable. Let $V(x)=\max_{k=1,\ldots,r}V_{k-1}(x)$. With Proposition 2.5 of \cite{lin1996smooth}, there exists a function $\beta$ of class $KL$ such that
\begin{flalign}
    \tilde V(x(t))
    \leq|m(t)|
    \leq\beta(\tilde V(x_0),t).
    \label{eq:cascade-decay-compo-tildeV}
\end{flalign}
Take
\begin{flalign}
    \underline\psi(s)=\inf_{|x|_{\C}\geq s}\tilde V(x),\;\;\bar\psi(s)=\sup_{|x|_{\C}\leq s}\tilde V(x),\;\;\forall s\geq0.
    \nonumber
\end{flalign}
Note that $V(x)$ is zero inside $\C$, positive for all $x\in\R^n\backslash\C$, and tends to infinity as $|x|_{\C}$ tends to infinity. Thus, $\underline\psi$ and $\bar\psi$ are continuous, non-decreasing and unbounded on $\R_{\geq0}$, and satisfy $\underline\psi(0)=\bar\psi(0)=0$. Choose two class $K_\infty$ functions $\underline\alpha$ and $\bar\alpha$ such that $\underline\alpha(s)\leq\underline\psi(s)/c$ and $\bar\alpha(s)\geq c\bar\psi(s)$ with $c>1$. Therefore,
\begin{flalign}
    \underline\alpha(|x|_{\C})\leq\underline\psi(|x|_{\C})\leq V(x) \leq\bar\psi(|x|_{\C})\leq\bar\alpha(|x|_{\C}).
    \nonumber
\end{flalign}
Then, with (\ref{eq:cascade-decay-compo-tildeV}),
\begin{flalign}
    |x(t)|_{\C}
    \leq\underline\alpha^{-1}(\beta(\bar\alpha(|x_0|_{\C}),t))
     \label{eq:cascade-asympt-stable}
\end{flalign}
which implies the asymptotical stability of system (\ref{eq:system-with-input}) with respect to $\C$.

The rest is to show the ISS of system (\ref{eq:system-with-input}) with respect to the set $\setS$. Let $V(x)=\max_{k=1,\ldots,r}V_{k-1}(x)$ with ${V}_{k-1}(x)=\max\{0,-\eta_{k-1}(x)\}$. Clearly, $\tilde V(x_0)\leq V(x_0)$ and
\begin{flalign}
    \tilde V(x(t))\geq V(x(t))-\max_{k=1,\ldots,r}\hat\alpha_k\circ\gamma(\|u\|)
    \nonumber
\end{flalign}
From (\ref{eq:cascade-decay-compo-tildeV}),
\begin{flalign}
    V(x(t))\leq\beta(V(x_0),t)+\max_{k=1,\ldots,r}\hat\alpha_k\circ\gamma(\|u\|)
    \nonumber
\end{flalign}
Similar to the derivation of (\ref{eq:cascade-asympt-stable}), there exists class $K_\infty$ functions $\underline\alpha$ and $\bar\alpha$ such that
\begin{flalign}\label{eq:cascade-decay-compo-x}
    |x|_{\setS}
    &\leq\underline\alpha\Big(\beta(\bar\alpha(|x_0|_{\setS}),t)
        +\max_{k=1,\ldots,r}\hat\alpha_k\circ\gamma(\|u\|)\Big)
    \nonumber\\
    &\leq\underline\alpha(2\beta(\bar\alpha(|x_0|_{\setS}),t))
        +\max_{k=1,\ldots,r}\underline\alpha(2\hat\alpha_k\circ\gamma(\|u\|)).
\end{flalign}
Thus, the ISS of system (\ref{eq:system-with-input}) with respect to $\setS$ follows.

\section*{Appendix III: Proof of Lemma \ref{lem:alpha-to-beta}}

\setcounter{subsection}{0}

From (\ref{eq:eta-geq-sub-eta-w}),
\begin{flalign}\label{eq:y-eq-sub-y-infw}
    \dot\eta(t)\geq-\alpha(\eta(t))+\alpha(\eta^*),\;\;\forall t\in[0,T).
\end{flalign}
Consider the comparison equation
\begin{flalign}\label{eq:y-eq-sub-y-w}
    \dot{y}=-\alpha(y)+\alpha(\eta^*),\;\;y(0)=\eta_0.
\end{flalign}

\begin{claim}\label{claim:compar-equi}
The comparison equation (\ref{eq:y-eq-sub-y-w}) has a unique solution $y(t)$ defined on $\R_{\geq0}$. Moreover,
\begin{flalign}
    y(t)=\beta(y_0-\eta^*,t)+\eta^*
    \label{eq:solution-compar-sys}
\end{flalign}
where $\beta:\R\times\R_{\geq0}\rightarrow\R$ is an extended class $KL$ function satisfying $\beta(s,0)=s$.
\end{claim}

Then the conclusion of Lemma \ref{lem:alpha-to-beta} follows, by applying Claim \ref{claim:compar-equi} and the standard comparison lemma \cite[Lem. 3.4]{khalil2002nonlinear} to (\ref{eq:y-eq-sub-y-infw}). Thus, the rest is to prove this claim.

\vskip5pt
\textbf{Proof of Claim 1.}
The local Lipschitzness of $\alpha$ implies that (\ref{eq:y-eq-sub-y-w}) has a unique solution $y(t)$ for each $y_0\in\R$. Since $y=\eta^*$ is an equilibrium point of (\ref{eq:y-eq-sub-y-w}) and $\dot y(t)<0$ (resp. $\dot{y}(t)>0$) when $y(t)>\alpha^{-1}(w)$ (resp. $y(t)<\alpha^{-1}(w)$), it follows that $-|y_0|\leq y(t) \leq |y_0|$. Therefore, the solution of (\ref{eq:y-eq-sub-y-w}) is bounded and can be extended indefinitely.

Take $\tilde{y}=y-\eta^*$, and then (\ref{eq:y-eq-sub-y-w}) can be rewritten as
\begin{flalign}\label{eq:diff-tilde-y}
    \dot{\tilde y}=-\hat\alpha(\tilde y),\;\;\tilde y(0)=y_0-\eta^*
\end{flalign}
where $\hat\alpha(s)=\alpha(s+\eta^*)-\alpha(\eta^*)$ with $\hat\alpha(0)=0$ is also a locally Lipschitz extended class $K_\infty$ function. Note that $\tilde y(t)\equiv0$ if $\tilde y_0=0$, since $\tilde y=0$ is an equilibrium of (\ref{eq:diff-tilde-y}). Without loss of generality, we assume $\tilde y_0\neq0$ in the following. By integration, the solution $\tilde y(t)$ of (\ref{eq:diff-tilde-y}) satisfies
\begin{flalign}\label{eq:init-prob-sol-form-A}
    -\int_{\tilde y(0)}^{\tilde y(t)}\dfrac{\text{d}r}{\hat\alpha(r)}=\int_{0}^{t}\text{d}\tau.
\end{flalign}
Define, for any $s\in\R\backslash\{0\}$,
\begin{flalign}\label{eq:init-prob-etadef}
    \eta(s)=
    \left\{
      \begin{array}{ll}
        -\int_{1}^{s}\frac{dr}{\hat\alpha(r)}, & \text{if } s>0 \\
        -\int_{-1}^{s}\frac{dr}{\hat\alpha(r)}, & \text{if } s<0
      \end{array}
    \right.
\end{flalign}
which is strictly decreasing on $(0,+\infty)$ and strictly increasing on $(-\infty,0)$. From the uniqueness of the solution of (\ref{eq:diff-tilde-y}), it follows that  $\tilde y(t)$ tends to zero if and only if $t$ tends to infinity, and thus, $\tilde y(t)\geq0$ (resp. $\tilde y(t)\leq0$) for all $t\geq0$ if $\tilde y_0\geq0$ (resp. $\tilde y_0\leq0$). Recalling (\ref{eq:init-prob-sol-form-A}) and (\ref{eq:init-prob-etadef}), the solution $\tilde y(t)$ of (\ref{eq:diff-tilde-y}) satisfies
\begin{flalign}
    \eta(\tilde y(t))-\eta(\tilde y(0))=t.
    \nonumber
\end{flalign}
Let
\begin{flalign}
    \beta(s,t)
    =\left\{
       \begin{array}{ll}
         \eta^{-1}(\eta(s)+t), & \text{if } s\neq0, \\
         0, & \text{if } s=0. \\
       \end{array}
     \right.
    \nonumber
\end{flalign}
Then $\tilde y(t)=\beta(\tilde y(0),t)$, and thus, (\ref{eq:solution-compar-sys}) holds for all $t\geq0$. The rest is to show that $\beta$ is of extended class $KL$. Since $\hat\alpha$ is locally Lipschitz, for each $s\in\R\backslash{0}$, $|\hat\alpha(s)|\leq K|s|$. Consequently,
\begin{flalign}
    &\lim_{s\rightarrow0^+}\eta(s)
    =\lim_{s\rightarrow0^+}\int_{s}^{1}\frac{dr}{\hat\alpha(r)}
    \geq\lim_{s\rightarrow0^+}\int_{s}^{1}\frac{dr}{Kr}
    =+\infty,
    \nonumber\\
    &\lim_{s\rightarrow0^-}\eta(s)
    =-\lim_{s\rightarrow0^-}\int_{-1}^{s}\frac{dr}{\hat\alpha(r)}
    \geq-\lim_{s\rightarrow0^-}\int_{-1}^{s}\frac{dr}{Kr}
    =+\infty.
     \nonumber
\end{flalign}
As a result,
\begin{flalign}
    \lim_{s\rightarrow+\infty}\eta^{-1}(s)=0.
    \nonumber
\end{flalign}
Since $\eta$ and $\eta^{-1}$ are continuous functions, $\beta$ is also continuous. For each fixed $t\geq0$,
\begin{flalign}
    \dfrac{\partial}{\partial s}\beta(s,t)=\frac{\eta'(s)}{\eta'(\beta(s,t))}=\frac{\hat\alpha\circ\eta^{-1}(\eta(s)+t)}{\hat\alpha(s)}>0,
     \nonumber
\end{flalign}
and thus, $\beta$ is strictly increasing on $s$. In addition,
\begin{flalign}
    \dfrac{\partial}{\partial t}\beta(s,t)=\frac{1}{\eta'(\beta(s,t))}=-\hat\alpha\circ\eta^{-1}(\eta(s)+t).
     \nonumber
\end{flalign}
Therefore, ${\partial\beta(s,t)}/{\partial t}<0$ for each fixed $s>0$ and ${\partial\beta(s,t)}/{\partial t}>0$ for each $s<0$. Because $\lim_{s\rightarrow+\infty}\eta^{-1}(s)=0$, $\beta(s,t)$ will decrease and increase to zero for each fixed $s>0$ and $s<0$, respectively, as $t$ tends to infinity.
\hfill $\Box$

\section*{Appendix IV: Proof of Theorem \ref{thm:non-sg-with-input}}

\setcounter{subsection}{0}

In order to prove Theorem \ref{thm:non-sg-with-input}, we introduce a useful inequality, that is, for any extended class $K_\infty$ functions $\gamma$ and $\sigma$, and any real numbers $a$ and $b$,
\begin{flalign}
    \gamma(a+b)\geq\min\{\gamma\circ(\Id+\sigma)(a),\gamma\circ(\Id+\sigma^{-1})(b)\}.
    \label{eq:weak-ieq-A}
\end{flalign}
This inequality is extended from \cite[Inequality (6)]{Jiang1994Small} by removing the positive definiteness assumption. It can be verified by combining the following two cases: i) if $b\geq\sigma(a)$, then $\gamma(a+b)\geq\gamma\circ(\Id+\sigma)(a)$; and ii) if $b\leq\sigma(a)$, then $\gamma(a+b)\geq\gamma\circ(\Id+\sigma^{-1})(b)$. Moreover, if $a,b\leq0$,
\begin{flalign}
    \gamma(a+b)\geq\gamma\circ(\Id+\sigma)(a)+\gamma\circ(\Id+\sigma^{-1})(b).
    \label{eq:weak-ieq-B}
\end{flalign}


\subsection{Proof of (i) of Theorem \ref{thm:non-sg-with-input}}\label{subsec:completeness-inter-con}

Suppose that, for any $T\in J(x_0,u)$, the solution $x(t)$ of system (\ref{eq:inter-sys-with-input}) is right maximally defined on $[0, T)$. Let
\begin{flalign}\label{eq:inter-conn-lya}
   V_{i,k-1}(x_i)=\max\{0,-\eta_{i,k-1}(x_i)\}
\end{flalign}
for $i=1,2$ and $k=1,\ldots,r$. From (\ref{eq:eta-nonlinear-sg}),
\begin{flalign}
    &\dot\eta_{i,k-1}(x_i(t))
     \nonumber\\
    &\;\;\;\;\geq-\alpha_{i,k}(\eta_{i,k-1}x_i(t))+\inf_{t\in[0,T)}\eta_{i,k}(x_i(t))
     \nonumber\\
    &\;\;\;\;\geq-\alpha_{i,k}(\eta_{i,k-1}x_i(t))-\sup_{t\in[0,T)}\max\{0,-\eta_{i,k}(x_i(t))\}.
     \label{eq:d-eta-alpha-supV}
\end{flalign}
By applying Lemma \ref{lem:alpha-to-beta} to (\ref{eq:d-eta-alpha-supV}) with taking $w(t)=-\|V_{i,k}\|_{[0,T)}=-\sup\nolimits_{t\in[0,T)}\max\{0,-\eta_{i,k}(t)\}$,
there exists an extended class $KL$ function $\rho_{i,k}$ satisfying $\rho_{i,k}(s,0)=s$ such that
\begin{flalign}
    &\eta_{i,k-1}(x_i(t))
     \nonumber\\
    &\;\;\;\;
      \geq\rho_{i,k}\Big(\eta_{i,k-1}(0)-\alpha_{i,k}^{-1}(-\|V_{i,k}\|_{[0,T)}),t\Big)
          +\alpha_{i,k}^{-1}(-\|V_{i,k}\|_{[0,T)})
     \nonumber\\
    &\;\;\;\;
      \geq\rho_{i,k}(\eta_{i,k-1}(0),t)+\alpha_{i,k}^{-1}(-\|V_{i,k}\|_{[0,T)})
     \nonumber\\
    &\;\;\;\;
      \geq\rho_{i,k}(-V_{i,k-1}(0),t)+\alpha_{i,k}^{-1}(-\|V_{i,k}\|_{[0,T)}),\;\forall t\in[0,T).
    \label{eq:rho-geq-rho-subV-alpha-V}
\end{flalign}
Let $\beta_{i,k}(s,t)=-\rho(-s,t)$. Clearly, $\beta_{i,k}$ is a class $KL$ function on $\R_{\geq0}\times\R_{\geq0}$. Because the term on the right-hand side of (\ref{eq:rho-geq-rho-subV-alpha-V}) is not positive, the combination of (\ref{eq:inter-conn-lya}) and (\ref{eq:rho-geq-rho-subV-alpha-V}) yields
\begin{flalign}\label{eq:vt-beta-alpha}
    V_{i,k-1}(x_i(t))
    &\leq\beta_{i,k}(V_{i,k-1}(0),t)
     \nonumber\\
    &\;\;\;\;\;\;\;\;\;\;
    -\alpha_{i,k}^{-1}(-\|V_{i,k-1}\|_{[0,T)}),\;\forall t\in[0,T).
\end{flalign}
Let $V(x)=\sum_{i=1}^2\sum_{k=1}^{r}V_{i,k-1}(x_i)$. Then,
\begin{flalign}
    \|V_{i,k-1}\|_{[0,T)}
    \leq\beta_{i,k}(V(0),0)-\alpha_{i,k}^{-1}(-\|V_{i,k-1}\|_{[0,T)}).
    \nonumber
\end{flalign}
With inequality (\ref{eq:weak-ieq-B}),
\begin{flalign}
    \|V_{i,0}\|_{[0, T)}
    &\leq\beta_{i,1}(V(0),0)
        -\alpha_{i,1}^{-1}(-\beta_{i,2}(V(0),0)+\alpha_{i,2}^{-1}(-\|V_{i,2}\|_{[0, T)}))
     \nonumber\\
    &\leq\beta_{i,1}(V(0),0)
        -\alpha_{i,1}^{-1}\circ(\Id+\sigma^{-1})(-\beta_{i,2}(V(0),0))
     \nonumber\\
    &\;\;\;\;\;\;\;\;\;\;\;\;
        -\alpha_{i,1}^{-1}\circ(\Id+\sigma)\circ\alpha_{i,2}^{-1}(-\|V_{i,2}\|_{[0, T)})
     \nonumber\\
    &\cdots
     \nonumber\\
    &\leq
     \hat\beta_{i,1}(V(0),0)
         -(\Id+\sigma)^{-1}\circ\hat\phi_{i,1}(-\|V_{3-i,0}\|_{[0, T)})
     \nonumber\\
    &\;\;\;\;\;\;\;\;\;\;\;\;
         -(\Id+\sigma)^{-1}\circ(-\Id)\circ\hat\gamma_{i,1}(\|u\|)
         \nonumber
\end{flalign}
where
\begin{flalign}
    &\hat\beta_{i,1}(s,t)
     \nonumber\\
    &=\beta_{i,1}(s,t)
        -\big[\alpha_{i,1}^{-1}\circ(\Id+\sigma^{-1})(-\beta_{i,2}(s,t))\big]
     \nonumber\\
    &\;\;\;\;
     -\big[\alpha_{i,1}^{-1}\circ(\Id+\sigma)\circ\alpha_{i,2}^{-1}\circ(\Id+\sigma^{-1})(-\beta_{i,3}(s,t))\big]
     \nonumber\\
    &\;\;\;\;
     -\cdots
     -\big[\alpha_{i,1}^{-1}\circ(\Id+\sigma)\circ\alpha_{i,2}^{-1}\circ\cdots\circ(\Id+\sigma)
      \nonumber\\
    &\;\;\;\;\;\;\;\;\;\;\;\;\;\;\;\;\;\;\;\;\;\;\;\;\;\;\;
     \circ\alpha_{i,r-1}^{-1}\circ(\Id+\sigma^{-1})(-\beta_{i,r}(s,t))\big].
     \nonumber
\end{flalign}
Thus,
\begin{flalign}
    \|V_{2,0}\|_{[0, T)}
    &\leq
     \hat\beta_{2,1}(V(0),0)
         -(\Id+\sigma)^{-1}\circ\hat\phi_{2,1}(-\|V_{1,0}\|_{[0, T)})
     \nonumber\\
    &\;\;\;\;
         -(\Id+\sigma)^{-1}\circ(-\Id)\circ\hat\gamma_{2,1}(\|u\|)
         \nonumber\\
    &\leq
     \hat\beta_{2,1}(V(0),0)
     -(\Id+\sigma)^{-1}\circ\hat\phi_{2,1}\Big(-\hat\beta_{1,1}(V(0),0)
     \nonumber\\
    &\;\;\;\;\;\;\;\;\;\;
     +(\Id+\sigma)^{-1}\circ\hat\phi_{1,1}(-\|V_{2,0}\|_{[0, T)})
     \nonumber\\
    &\;\;\;\;\;\;\;\;\;\;
      +(\Id+\sigma)^{-1}(-\hat\gamma_{1,1}(\|u\|))\Big)
      \nonumber\\
    &\;\;\;\;
         -(\Id+\sigma)^{-1}(-\hat\gamma_{2,1}(\|u\|))
     \nonumber\\
    &\leq
     \hat\beta_{2,1}(V(0),0)
     \nonumber\\
    &\;\;\;\;
      -(\Id+\sigma)^{-1}\circ\hat\phi_{2,1}
      \circ(\Id+\sigma^{-1})\Big(-\hat\beta_{1,1}(V(0),0)
      \nonumber\\
    &\;\;\;\;\;\;\;\;\;\;
         +(\Id+\sigma)^{-1}(-\hat\gamma_{1,1}(\|u\|))\Big)
     \nonumber\\
    &\;\;\;\;
     -(\Id+\sigma)^{-1}\circ\hat\phi_{2,1}\circ\hat\phi_{1,1}(-\|V_{2,0}\|_{[0, T)})
     \nonumber\\
    &\;\;\;\;
     -(\Id+\sigma)^{-1}(-\hat\gamma_{2,1}(\|u\|)).
     \nonumber
\end{flalign}
With the small-gain condition (\ref{eq:sg-cond}) and the equality
\begin{flalign}
    \Big(\Id+(\Id+\sigma)^{-1}\circ(-\Id)\Big)^{-1}(s)
        =-(\Id+\sigma^{-1})\circ(-\Id)(s),
    \nonumber
\end{flalign}
we have
\begin{flalign}\label{eq:V2-sup}
    \|V_{2,0}\|_{[0, T)}
    &\leq
     \big[\hat\beta_{2,1}(V(0),0)
     -(\Id+\sigma)^{-1}\circ\hat\phi_{2,1}\circ(\Id+\sigma^{-1})^2
     \nonumber\\
    &\;\;\;\;\;\;\;\;\;\;
     \circ(-\Id)\circ\hat\beta_{1,1}(V(0),0)\big]
     \nonumber\\
    &\;\;\;\;
     -[(\Id+\sigma^{-1})\circ\hat\phi_{2,1}\circ(\Id+\sigma^{-1})(-\hat\gamma_{1,1}(\|u\|))
     \nonumber\\
    &\;\;\;\;\;\;\;\;\;\;
     +(\Id+\sigma)^{-1}(-\hat\gamma_{2,1}(\|u\|))]
     \nonumber\\
    &\;\;\;\;
     -(\Id+\sigma)^{-1}(-\|V_{2,0}\|_{[0, T)})
     \nonumber\\
    &\leq
     -(\Id+\sigma^{-1})\Big(-\hat\beta_{2,1}(V(0),0)
     \nonumber\\
    &\;\;\;\;\;\;\;\;\;\;
     +(\Id+\sigma)^{-1}\circ\hat\phi_{2,1}
     \circ(\Id+\sigma^{-1})^2(-\hat\beta_{1,1}(V(0),0))
     \nonumber\\
    &\;\;\;\;\;\;\;\;\;\;
     +(\Id+\sigma)^{-1}\circ\hat\phi_{2,1}
      \circ(\Id+\sigma^{-1})(-\hat\gamma_{1,1}(\|u\|))
     \nonumber\\
    &\;\;\;\;\;\;\;\;\;\;
     +(\Id+\sigma)^{-1}(-\hat\gamma_{2,1}(\|u\|))
    \Big)
     \nonumber\\
    &\leq\delta_{2,1}(V(0))+\Delta_{2,1}(\|u\|)
\end{flalign}
where $\delta_{2,1}$ and $\Delta_{2,1}$ are class $K_\infty$ functions defined as
\begin{flalign}
    &\delta_{2,1}(s)
    =-(\Id+\sigma^{-1})^2\Big(-\hat\beta_{2,1}(V(0),0)
     \nonumber\\
    &\;\;\;\;\;\;\;\;\;\;\;\;\;\;\;\;\;\;
     +(\Id+\sigma)^{-1}\circ\phi_{2,1}\circ(\Id+\sigma^{-1})^2(-\beta_{1,1}(V(0),0))\Big),
     \nonumber\\
    &\Delta_{2,1}(s)=-(\Id+\sigma^{-1})\circ(\Id+\sigma)\Big((\Id+\sigma)^{-1}(-\hat\gamma_{2,1}(s))
     \nonumber\\
    &\;\;\;\;\;\;\;\;\;\;\;\;\;\;\;\;\;\;\;\;\;\;\;\;
    +(\Id+\sigma)^{-1}\circ\hat\phi_{2,1}\circ(\Id+\sigma^{-1})(-\hat\gamma_{1,1}(s))\Big).
     \nonumber
\end{flalign}
Due to the symmetry between $V_{1,0}$ and $V_{2,0}$, we construct functions $\delta_{1,1}$ and $\Delta_{1,1}$ of class $K_\infty$ such that
\begin{flalign}\label{eq:up-bound-V10}
    \|V_{1,0}\|_{[0, T)}\leq\delta_{1,1}(V(0))+\Delta_{1,1}(\|u\|).
\end{flalign}
Because $T$ is arbitrary on $J(x_0,u)$ and the right-hand sides of (\ref{eq:V2-sup}) and (\ref{eq:up-bound-V10}) are independent of $T$, we have
\begin{flalign}
    \|V_{i,0}\|_{J(x_0,u)}\leq\delta_{i,1}(V(0))+\Delta_{i,1}(\|u\|),\;\;i=1,2.
    \nonumber
\end{flalign}
Since $V_{i,0}$ is zero in the set $\setS_{i,0}$ and positive in $\R^n\backslash\setS_{i,0}$, with a similar argument of (\ref{eq:cascade-decay-compo-x}), there exists a class $K_\infty$ function $\underline\alpha_{i,1}$ such that
\begin{flalign}
    |x(t)|_{\setS_{i,0}}
    &\leq\underline\alpha_{i,1}^{-1}(\|V_{i,0}\|_{J(x_0,u)})
     \nonumber\\
    &\leq\underline\alpha_{i,1}^{-1}\big(\delta_{i,1}(V(0))+\Delta_{i,1}(\|u\|)\big),\;\;\forall t\in J(x_0,u).
    \nonumber
\end{flalign}
Thus, the distance from $x(t)$ to $\setS_{1,0}\bigcap\setS_{2,0}$ is finite. Together with boundness of the distance between $x(t)$ and $\R^n\backslash(\setS_{1,0}\bigcap\setS_{2,0})$, the solution $x(t)$ exists for all $t\in I(x_0,u)=J(x_0,u)$.

\subsection{Proof of (ii) of Theorem \ref{thm:non-sg-with-input}}\label{sec:proof-invariant}

With the existence of solutions on $I(x_0,u)=J(x_0,u)$, $\inf_{t\in I(x_0,u)}\eta_{i,k}(t)$ is well defined for $i=1,2$ and $k=1,\ldots,r$, and thus, according to (\ref{eq:eta-nonlinear-sg}),
\begin{flalign}
    \dot\eta_{i,k-1}(x_i(t))
    \geq-\alpha_{i,k}(\eta_{i,k-1}x_i(t))+\inf_{t\in I(x_0,u)}\eta_{i,k}(x_i(t)).
    \nonumber
\end{flalign}
With Lemma \ref{lem:alpha-to-beta} (taking $w(t)=\inf_{t\in I(x_0,u)}\eta_{i,k}(t)$), there exists an extended class $KL$ function $\rho_{i,k}$ satisfying $\rho_{i,k}(s,0)=s$ such that
\begin{flalign}
    \eta_{i,k-1}(t)\geq\rho_{i,k}(\eta_{i,k-1}(0)-\eta_{i,k-1}^*,t)+\eta_{i,k-1}^*,\;\;\forall t\in I(x_0,u)
    \nonumber
\end{flalign}
where $\eta_{i,k-1}^*=\alpha_{i,k}^{-1}(\inf_{t\in I(x_0,u)}\eta_{i,k}(t))$. Since the mapping $t\mapsto\rho_{i,k}(s,t)$ is strictly increasing (resp. decreasing) for each $s<0$ (resp. $s>0$),
\begin{flalign}\label{eq:inf-eta}
    \eta_{i,k-1}(t)
    &\geq\min\{\rho_{i,k}(\eta_{i,k-1}(0)-\eta_{i,k}^*,0)+\eta_{i,k-1}^*, \eta_{i,k-1}^*\}
     \nonumber\\
    &\geq\min\{\eta_{i,k-1}(0), \eta_{i,k-1}^*\}
     \nonumber\\
    &\geq\min\{\hat\phi_{i,k}(-\hat\gamma_{3-i,1}(\|u\|)),
    -\hat\gamma_{i,k}(\|u\|),
    \alpha_{i,k}^{-1}(\inf_{t\in I(x_0,u)}\eta_{i,k}(t))\}
     \nonumber\\
    &\geq\min\{\hat\phi_{i,k}(-\hat\gamma_{3-i,1}(\|u\|)),
    -\hat\gamma_{i,k}(\|u\|),
     \nonumber\\
    &\;\;\;\;\;\;\;\;\;\;\;\;\;\;\;\;\;\;\;\;\;\;
    (\Id+\sigma)\circ\alpha_{i,k}^{-1}(\inf_{t\in I(x_0,u)}\eta_{i,k}(t))\}
\end{flalign}
for each  $x(0)\in\C$ and any $\sigma$ of extended class $K_\infty$. Because $\hat\phi_{i,k}(s)=(\Id+\sigma)\circ\alpha_{i,k}^{-1}\circ\hat\phi_{i,k+1}(s)$ and $\hat\gamma_{i,k}(s)=-(\Id+\sigma)\circ\alpha_{i,k}^{-1}(-\hat\gamma_{i,k+1}(s))$, we have
\begin{flalign}\label{eq-estim-hi-A}
    h_i(t)
    &=\eta_{i,0}(t)
     \nonumber\\
    &\geq\min\{\hat\phi_{i,1}(-\hat\gamma_{3-i,1}(\|u\|)),-\hat\gamma_{i,1}(\|u\|),
     \nonumber\\
    &\;\;\;\;\;\;\;\;\;\;\;\;\;\;\;\;\;\;\;\;\;\;
        (\Id+\sigma)\circ\alpha_{i,1}^{-1}(\inf_{t\in I(x_0,u)}\eta_{i,1}(t))\}
     \nonumber\\
    &\geq\min\{\hat\phi_{i,1}(-\hat\gamma_{3-i,1}(\|u\|)),-\hat\gamma_{i,1}(\|u\|),
     \nonumber\\
    &\;\;\;\;\;\;\;\;\;
     (\Id+\sigma)\circ\alpha_{i,1}^{-1}\circ(\Id+\sigma)\circ\alpha_{i,2}^{-1}(\inf_{t\in I(x_0,u)}\eta_{i,2}(t))\}
     \nonumber\\
    &\;\;\;\;\;\;\;\;\;\;\cdots
     \nonumber\\
    &\geq\min\{\hat\phi_{i,1}(-\hat\gamma_{3-i,1}(\|u\|)),-\hat\gamma_{i,1}(\|u\|),
     \nonumber\\
    &\;\;\;\;\;\;\;\;\;
     (\Id+\sigma)\circ\alpha_{i,1}^{-1}\circ\cdots\circ(\Id+\sigma)\circ\alpha_{i,r}^{-1}(\inf_{t\in I(x_0,u)}\eta_{i,r}(t))\}
\end{flalign}
for all $t\in I(x_0,u)$. In addition, by combining (\ref{eq:issf-sg-externalinput}) and (\ref{eq:weak-ieq-A}),
\begin{flalign}\label{eq:issf-sg-externalinput-rewrite}
    \eta_{i,r}(t)
    &\geq\min\{(\Id+\sigma)\circ\phi_i(\inf_{t\in I(x_0,u)}h_{3-i}(t)),
     \nonumber\\
    &\;\;\;\;\;\;\;\;\;\;\;\;\;\;\;\;\;\;\;\;\;\;
    (\Id+\sigma^{-1})(-\gamma_i(\|u\|))\},\;\;\forall t\in I(x_0,u)
\end{flalign}
Substituting (\ref{eq:issf-sg-externalinput-rewrite}) into (\ref{eq-estim-hi-A}),
\begin{flalign}
    h_i(t)
    &\geq\min\{\hat\phi_{i,1}(-\hat\gamma_{3-i,1}(\|u\|)),
     \nonumber\\
    &\;\;\;\;\;\;\;\;\;\;\;\;\;\;\;\;\;\;
    -\hat\gamma_{i,1}(\|u\|),
    \hat\phi_{i,1}(\inf_{t\in I(x_0,u)}h_{3-i}(t))\},\;\;\forall t\in I(x_0,u)
    \nonumber
\end{flalign}
which implies
\begin{subequations}\label{eq:inf-hi}
\begin{flalign}
    &\inf_{t\in I(x_0,u)}h_1(t)
     \geq\min\{\hat\phi_{1,1}(-\hat\gamma_{2,1}(\|u\|)),
     \nonumber\\
    &\;\;\;\;\;\;\;\;\;\;\;\;\;\;\;\;\;\;\;\;\;\;\;\;\;\;\;\;\;\;
      -\hat\gamma_{1,1}(\|u\|),\hat\phi_{1,1}(\inf_{t\in I(x_0,u)}h_{2}(t))\},
     \label{eq:inf-hi-A}\\
    &\inf_{t\in I(x_0,u)}h_2(t)
     \geq\min\{\hat\phi_{2,1}(-\hat\gamma_{1,1}(\|u\|)),
     \nonumber\\
    &\;\;\;\;\;\;\;\;\;\;\;\;\;\;\;\;\;\;\;\;\;\;\;\;\;\;\;\;\;\;
     -\hat\gamma_{2,1}(\|u\|),\hat\phi_{2,1}(\inf_{t\in I(x_0,u)}h_{1}(t))\}.
     \label{eq:inf-hi-B}
\end{flalign}
\end{subequations}
By substituting (\ref{eq:inf-hi-B}) into (\ref{eq:inf-hi-A}),
\begin{flalign}
    &\inf_{t\in I(x_0,u)}h_2(t)
     \nonumber\\
    &\;\;\;\;
     \geq\min\{\hat\phi_{2,1}(-\hat\gamma_{1,1}(\|u\|)),-\hat\gamma_{2,1}(\|u\|),\hat\phi_{2,1}(\inf_{t\in I(x_0,u)}h_{1}(t))\}
     \nonumber\\
    &\;\;\;\;
     \geq\min\{\hat\phi_{2,1}(-\hat\gamma_{1,1}(\|u\|)),-\hat\gamma_{2,1}(\|u\|),
     \nonumber\\
    &\;\;\;\;\;\;\;\;\;\;\;\;\;\;\;
            \hat\phi_{2,1}\circ\hat\phi_{1,1}(-\hat\gamma_{2,1}(\|u\|)), \hat\phi_{2,1}\circ\hat\phi_{1,1}(\inf_{t\in I(x_0,u)}h_2(t))\}
     \nonumber\\
    &\;\;\;\;
     \geq\min\{\hat\phi_{2,1}(-\hat\gamma_{1,1}(\|u\|)),
        -\hat\gamma_{2,1}(\|u\|), \hat\phi_{2,1}\circ\hat\phi_{1,1}(\inf_{t\in I(x_0,u)}h_2(t))\}
     \nonumber
\end{flalign}
where the third inequality results from the small-gain condition (\ref{eq:sg-cond}). If $\inf_{t\in I(x_0,u)}h_2(t)\geq0$,
\begin{flalign}
    \hat\phi_{2,1}\circ\hat\phi_{1,1}(\inf_{t\in I(x_0,u)}h_2(t))\geq\min\{\hat\phi_{2,1}(-\hat\gamma_{1,1}(\|u\|)),-\hat\gamma_{2,1}(\|u\|)\},
    \nonumber
\end{flalign}
and if $\inf_{t\in I(x_0,u)}h_2(t)\leq0$, then, using the small-gain condition (\ref{eq:sg-cond}) again,
\begin{flalign}
    \hat\phi_{2,1}\circ\hat\phi_{1,1}(\inf_{t\in I(x_0,u)}h_2(t))\geq\inf_{t\in I(x_0,u)}h_2(t).
    \nonumber
\end{flalign}
Thus,
\begin{flalign}\label{eq:inf-hi-A-final}
    \inf_{t\in I(x_0,u)}h_2(t)
     \geq\min\{\hat\phi_{2,1}(-\hat\gamma_{1,1}(\|u\|)),
        -\hat\gamma_{2,1}(\|u\|)\}.
\end{flalign}
Substituting (\ref{eq:inf-hi-A-final}) into (\ref{eq:issf-sg-externalinput-rewrite}),
\begin{flalign}
    \inf_{t\in I(x_0,u)}\eta_{1,r}(t)
    &\geq\min\{(\Id+\sigma)\circ\phi_1\circ\hat\phi_{2,1}(-\hat\gamma_{1,1}(\|u\|)),
     \nonumber\\
    &\;\;\;\;\;
    (\Id+\sigma)\circ\phi_1(-\hat\gamma_{2,1}(\|u\|)),
    (\Id+\sigma^{-1})(-\gamma_1(\|u\|))\}.
    \nonumber
\end{flalign}
Because
\begin{flalign}
    &(\Id+\sigma)\circ\phi_1\circ\hat\phi_{2,1}(-\hat\gamma_{1,1}(\|u\|))
     \nonumber\\
    &\;\;\;\;
     =(\Id+\sigma)\circ\phi_1\circ\hat\phi_{2,1}\circ(\Id+\sigma)\circ\alpha_{1,1}^{-1}
     \nonumber\\
    &\;\;\;\;\;\;\;\;\;\;\;\;
      \circ\cdots\circ(\Id+\sigma)\circ\alpha_{1,r}^{-1}\circ(\Id+\sigma^{-1})(-\gamma_1(\|u\|))
     \nonumber\\
    &\;\;\;\;
     =(\Id+\sigma)\circ\phi_1\circ\hat\phi_{2,1}\circ\hat\phi_{1,1}
     \nonumber\\
    &\;\;\;\;\;\;\;\;\;\;\;\;
    \circ\phi_1^{-1}\circ(\Id+\sigma)^{-1}\circ(\Id+\sigma^{-1})(-\gamma_1(\|u\|))
     \nonumber\\
    &\;\;\;\;
     \geq(\Id+\sigma^{-1})(-\gamma_1(\|u\|)),
     \nonumber
\end{flalign}
we have
\begin{flalign}\label{eq:inf-eta-ir-one}
    \inf_{t\in I(x_0,u)}\eta_{1,r}(t)
    &\geq\min\{
     (\Id+\sigma)\circ\phi_1(-\hat\gamma_{2,1}(\|u\|)),
     \nonumber\\
    &\;\;\;\;\;\;\;\;\;\;\;\;\;\;\;\;\;\;\;\;\;\;\;\;\;\;\;\;
    (\Id+\sigma^{-1})(-\gamma_1(\|u\|))\}.
\end{flalign}
Thus,
\begin{flalign}\label{eq:inf-eta-ir-bound}
    \eta_{1,r}(x_i(t))\geq\inf_{t\in I(x_0,u)}\eta_{1,r}(t)\geq -v_1,\;\;\forall t\in I(x_0,u)
\end{flalign}
where
\begin{flalign}
    v_1=-\min\{(\Id+\sigma)\circ\phi_1(-\hat\gamma_{2,1}(\|u\|)),(\Id+\sigma^{-1})(-\gamma_{1}(\|u\|))\}.
    \nonumber
\end{flalign}
Thus, by recalling (\ref{eq:eta-nonlinear-sg}) and then using Theorem \ref{thm:HOISSF-BF}, we obtain that $x_1(t)$ does not leave the set $\bigcap_{k=1}^{r}\C_{1,k-1}$ for all $t\in I(x_0,u)$. Similarly, $x_2(t)$ always stays in the set $\bigcap_{k=1}^{r}\C_{2,k-1}$ as well. Thus, $\C=\bigcap_{i=1,2}\bigcap_{k=1}^{r}\C_{i,k-1}$ is robustly forward invariant. Because $\C$ is a subset of $\C_{1,0}\bigcap\C_{2,0}$, $x(t)$ always stays inside $\C_{1,0}\bigcap\C_{2,0}$ if $x_0\in\C$, and thus, system (\ref{thm:non-sg-with-input}) is ISSf on $\setS_{1,0}\bigcap\setS_{2,0}$.

\subsection{Proof of (iii) of Theorem \ref{thm:non-sg-with-input}}\label{sec:proof-invariant-set-ISS}

From (i) of Theorem \ref{thm:non-sg-with-input}, $J(x_0,u)=\R_{\geq0}$ implies that system (\ref{eq:inter-sys-with-input}) is forward complete.

Now we consider the Lyapunov function candidate in (\ref{eq:inter-conn-lya}). With a similar derivation of (\ref{eq:V2-sup}), we can construct functions $\delta_{i,k}$ and $\Delta_{i,k}$ of class $K_\infty$ such that
\begin{flalign}
    \|V_{i,k-1}\|\leq\delta_{i,k}(V(0))+\Delta_{i,k}(\|u\|)
    \nonumber
\end{flalign}
for $i=1,2$ and $k=2,\ldots,r$. Then, with (\ref{eq:V2-sup}) and (\ref{eq:up-bound-V10}),
\begin{flalign}\label{eq:bound-of-V}
    \|V\|
    =\sum_{i=1}^2\sum_{k=1}^r\|V_{i,k-1}\|
    \leq\delta(V(0))+\Delta(\|u\|)
\end{flalign}
where $\delta(s)=\sum_{i=1}^{2}\sum_{k=1}^r\delta_{i,k}(s)$ and $\Delta(s)=\sum_{i=1}^{2}\sum_{k=1}^r\Delta_{i,k}(s)$.
On the other hand, (\ref{eq:vt-beta-alpha}) can be rewritten as
\begin{subequations}\label{eq:V-iksubone-beta-alpha}
\begin{flalign}
    &V_{i,k-1}(t_{k-1}^1)
     \leq\beta_{i,k}(V_{i,k-1}(t_{k-1}^0),t_{k-1}^1-t_{k-1}^0)
     \nonumber\\
    &\;\;\;\;\;\;\;\;\;\;\;\;\;\;\;\;\;\;\;\;
         -\alpha_{i,k}^{-1}(-\|V_{i,k}\|_{[t_{k-1}^0,t_{k-1}^1]}),\;\;k=1,\ldots,r-1,
     \\
    &V_{i,r-1}(t_{r-1}^1)
     \leq\beta_{i,r}(V_{i,r-1}(t_{r-1}^0),t_{r-1}^1-t_{r-1}^0)
     \nonumber\\
    &\;\;\;\;\;\;\;\;\;\;\;\;\;\;\;\;\;\;\;\;
         -\alpha_{i,r}^{-1}(\phi_i(-\|V_{3-i,0}\|_{[t_{r-1}^0,t_{r-1}^1]})-\gamma_i(\|u_i\|)).
\end{flalign}
\end{subequations}
where $t_{k-1}^1\geq t_{k-1}^0 \geq0$. For any $t\geq0$, take
\begin{flalign}
    &t_{2,0}^0=\frac{2r}{2r+1}t,t_{2,1}^0=\frac{2r-1}{2r+1}t,\ldots,t_{2,r-1}^0=\frac{r+1}{2r+1}t,
     \nonumber\\
    &t_{1,0}^0=\frac{r}{2r+1}t,t_{1,1}^0=\frac{r-1}{2r+1}t,\ldots,t_{1,r-1}=\frac{1}{2r+1}t
     \nonumber
\end{flalign}
and
\begin{flalign}
    &t_{2,0}^1=t,t_{2,1}^1\in[t_{2,0}^0,t],\ldots,t_{2,r-1}^1\in[t_{2,r-2}^0,t],
     \nonumber\\
    &t_{1,0}^1\in[t_{2,r-1}^0,t],t_{1,1}^1\in[t_{1,0}^0,t],\ldots,t_{1,r-1}^1\in[t_{1,r-2}^0,t].
     \nonumber
\end{flalign}
Clearly, $t_{i,k}^1-t_{i,k}^0\geq t/(2r+1)$. For notational convenience, let $\omega=1/(2r+1)$. Then the combination of (\ref{eq:bound-of-V}) and (\ref{eq:V-iksubone-beta-alpha}) yields
\begin{flalign}
    &V_{i,k-1}(t_{i,k-1}^1)
     \leq\beta_{i,k}(s_\infty,\omega t)
     \nonumber\\
    &\;\;\;\;\;\;\;\;\;\;\;\;\;\;\;\;\;\;\;\;\;\;
         -\alpha_{i,k}^{-1}(-\|V_{i,k}\|_{[t_{i,k-1}^0,t]}),\;\;k=1,\ldots,r-1,
     \nonumber\\
    &V_{i,r-1}(t_{i,r-1}^1)
     \leq\beta_{i,r}(s_\infty,\omega t)
     \nonumber\\
    &\;\;\;\;\;\;\;\;\;\;\;\;\;\;\;\;\;\;\;\;\;\;
         -\alpha_{i,r}^{-1}(\phi_i(-\|V_{3-i,0}\|_{[t_{3-i,0}^0,t]})-\gamma_i(\|u_i\|))
     \nonumber
\end{flalign}
where $s_\infty:=\delta(V(0))+\Delta(\|u\|)$. Using a similar derivation of (\ref{eq:V2-sup}),
\begin{flalign}
    V_{2,0}(t)
    &\leq
     \big[\hat\beta_{2,1}(s_\infty,\omega t)
    -(\Id+\sigma)^{-1}\circ\hat\phi_{2,1}
     \nonumber\\
    &\;\;\;\;\;\;\;\;
    \circ(\Id+\sigma^{-1})^2\circ(-\Id)\circ\hat\beta_{1,1}(s_\infty,\omega t)\big]
     \nonumber\\
    &\;\;\;\;
     -(\Id+\sigma)^{-1}(-\|V_{2,0}\|_{[\omega t,\infty)})
     \nonumber\\
    &\;\;\;\;
     -\big[(\Id+\sigma^{-1})\circ\hat\phi_{2,1}\circ(\Id+\sigma^{-1})(-\hat\gamma_{1,1}(\|u\|))
      \nonumber\\
     &\;\;\;\;\;\;\;\;
     +(\Id+\sigma)^{-1}(-\hat\gamma_{2,1}(\|u\|))\big].
\end{flalign}
Note that $0<\omega<1$ and $\Id+(\Id+\sigma^{-1})(-s)=-(\Id+\sigma^{-1})(-s)$ is of class $K_\infty$ on $\R_{\geq0}$. According to \cite[Lemma A.1]{Jiang1994Small} (by taking $\lambda(s)=(\Id+\sigma)(s)$ and $\rho(s)=-(\Id+\sigma)^{-1}(-s)$), there exists a class $KL$ function $\varrho_{2,1}$ such that
\begin{flalign}
    V_{2,0}(t)\leq\varrho_{2,1}(s_\infty,t)+\Delta_{2,1}(\|u\|).
    \nonumber
\end{flalign}
Analogously,
\begin{flalign}
    V_{i,k-1}(t)\leq\varrho_{i,k}(s_\infty,t)+\Delta_{i,k}(\|u\|)
    \nonumber
\end{flalign}
for $i=1,2$ and $k=1,\ldots,r$, where $\varrho_{i,k}$ is of class $KL$. Since the mapping $s\mapsto\varrho_{i,k}(s,t)$ is increasing,
\begin{flalign}
    V_{i,k-1}(t)
    &\leq\varrho_{i,k}(\delta(V(0))+\Delta(\|u\|),t)+\Delta_{i,k}(\|u\|)
     \nonumber\\
    &
     \leq\varrho_{i,k}(2\delta(V(0)),t)
         +\varrho_{i,k}(2\Delta(\|u\|),0)
         +\Delta_{i,k}(\|u\|)
     \nonumber
\end{flalign}
Thus,
\begin{flalign}\label{eq:cascade-decay-compo-V}
    V(t)
    &\leq\sum_{i=1}^2\sum_{k=1}^r\varrho_{i,k}\Big(2\delta(V(0)),t\Big)
        \nonumber\\
    &\;\;\;\;
         +\Delta(\|u\|)
         +\sum_{i=1}^2\sum_{k=1}^r\varrho_{i,k}\Big(2\Delta(\|u\|),0\Big).
\end{flalign}
Then the conclusion follows with the same argument of (\ref{eq:cascade-decay-compo-x}).


\ifCLASSOPTIONcaptionsoff
  \newpage
\fi


\end{document}